%% file: 1SCMv5.1online.tex
\documentclass[12pt,preprint]{aastex}

\newcommand{\msun}{${\rm M}_\odot$}

\newcommand{\dimho}{km~s$^{-1}$~Mpc$^{-1}$}
\newcommand{\kmpsec}{km~s$^{-1}$}
\renewcommand{\nodata}{ ~$\cdots$~ }
\renewcommand{\mark}{$^*$~}
\newcommand{\nad}{\ion{Na}{1}~D}
\shorttitle{Standardized Candle Method for SNe IIP}
\shortauthors{Olivares E. et al.}

\begin{document}

\title{The Standardized Candle Method for \\ Type II Plateau
  Supernovae}

\author{Felipe Olivares E.\altaffilmark{1,2}, 
        Mario Hamuy\altaffilmark{1}, 
        Giuliano Pignata\altaffilmark{1,3}, 
        Jos\'e Maza\altaffilmark{1}, 
        Melina Bersten\altaffilmark{1}, 
        Mark M. Phillips\altaffilmark{4}, 
        Nicholas B. Suntzeff\altaffilmark{5}, 
        Alexei V. Filippenko\altaffilmark{6}, 
        Nidia I. Morrel\altaffilmark{4}, 
        Robert P. Kirshner\altaffilmark{7}, and 
        Thomas Matheson\altaffilmark{8}}

\altaffiltext{1}{Departamento de Astronom\'ia, Universidad de Chile,
  Santiago, Chile.}

\altaffiltext{2}{Max-Planck-Institut f\"ur Extraterrestrische Physik,
  85740 Garching, Germany;\\
  email: \anchor{mailto:felipe@mpe.mpg.de}{felipe@mpe.mpg.de}.}

\altaffiltext{3}{Departamento de Ciencias Fisicas, Universidad Andres
  Bello, Avda. Republica 252, Santiago, Chile.}

\altaffiltext{4}{Las Campanas Observatory, Carnegie Observatories,
  Casilla 601, La Serena, Chile.}

\altaffiltext{5}{Texas A\&M University, Physics Department, College
  Station, TX 77843.}

\altaffiltext{6}{Department of Astronomy, University of California,
  Berkeley, CA 94720--3411.}

\altaffiltext{7}{Harvard-Smithsonian Center for Astrophysics, 60
  Garden Street, Cambridge, MA, 02138.}

\altaffiltext{8}{National Optical Astronomy Observatory, 950 North
  Cherry Avenue, Tucson, AZ 85719-4933.}

\setcounter{footnote}{8}


\include{aAbst}

\input{c1Intr}
\input{c2Data}

\input{c3Meth0}

\input{c4Anly0}

\input{c5Disc}

\input{c6Conc}

\input{zAckn}
\input{zFaci}


\clearpage

\include{0_online/f31LCfit}

\include{0_online/f32Colr}

\include{0_online/f33Vels}

\include{0_online/f34Spec}
\include{0_online/f35VIVR}

\include{0_online/f36VIBV}

\include{0_online/f401AviAspc}

\include{0_online/f402AviAnad}
\include{0_online/f403LmVe}

\include{0_online/f404HDbA}
\include{0_online/f405HDvA}

\include{0_online/f406HDiA}

\include{0_online/f407HDbviRv}

\include{0_online/f408RvBeta}

\include{0_online/f409resM}

\include{0_online/f410resCor}

\include{0_online/f410EPMSCM}
\include{0_online/f411EPMSCM}

\include{t21Teles}
\include{t22List}
\include{t31Pars}
\include{t32Exts}
\include{t41HDs}
\include{t42Param}
\include{t43Calib}

\include{t44Dist}
\include{t45Comp}

\end{document}

%% file: aAbst.tex
\begin{abstract}
  In this paper we study the ``standardized candle method'' using a
  sample of 37 nearby (redshift $z<0.06$) Type II plateau supernovae
  having $BVRI$ photometry and optical spectroscopy.  An analytic
  procedure is implemented to fit light curves, color curves, and
  velocity curves. We find that the $V-I$ color toward the end of the
  plateau can be used to estimate the host-galaxy reddening with a
  precision of $\sigma(A_V) = 0.2$ mag. The correlation between
  plateau luminosity and expansion velocity previously reported in the
  literature is recovered. Using this relation and assuming a standard
  reddening law ($R_V=3.1$), we obtain Hubble diagrams in the $BVI$
  bands with dispersions of $\sim$~0.4 mag. Allowing $R_V$ to vary and
  minimizing the spread in the Hubble diagrams, we obtain a dispersion
  range of 0.25--0.30 mag, which implies that these objects can
  deliver relative distances with precisions of 12--14\%. The
  resulting best-fit value of $R_V$ is $1.4 \pm 0.1$.
\end{abstract}


%% file: c1Intr.tex
\section{INTRODUCTION} \label{INTR}

Core-collapse SNe (CCSNe) are in general closely associated with
star-forming regions in late-type galaxies \citep[e.g.,][]{AnJ08}.
Therefore, they have been attributed to stars born with initial mass
$>$~8~\msun\ that undergo the collapse of their iron cores after a few
million years of evolution, and the subsequent ejection of their
envelopes \citep[e.g.,][]{Bur00}. According to theoretical studies,
these SNe leave a compact object as a remnant, either a neutron star
or a black hole \citep{BaZ34,Arn96,Jan07}.

Among CCSNe, we can observationally distinguish those with prominent
hydrogen lines in their spectra \citep[dubbed Type~II;][]{Min41},
those with no H but strong He lines (Type~Ib), and those lacking H or
He lines (Type~Ic); see \citep{Fil97} and references therein. Although
all of these objects are thought to share largely the same explosion
mechanism, their different observational properties are explained in
terms of how much of their H-rich and He-rich envelopes were retained
prior to explosion. When the star explodes with a significant fraction
of its initial H-rich envelope intact, it should display a H-rich
spectrum and a light curve characterized by an optically thick phase
of $\sim$~100~days of nearly constant luminosity called a ``plateau,''
which comes to an end at the point when all of the hydrogen in the
envelope has recombined. This phase is followed by a sudden drop
of~2--3 mag and thereafter by an exponential decay in luminosity
caused by the radioactive decay of $^{56}$Co into $^{56}$Fe
\citep[e.g.,][]{Nad03,Utr07}. Nearly~50\% of all CCSNe belong to this
class of ``Type~II plateau'' SNe \citep[SNe~IIP;][]{Cap99,Bot08}.

A decade ago, the standardized luminosities of Type~Ia SNe
\citep[e.g.,][]{Phi93,Ha96a,RPK96,Phi99} led to the construction of
Hubble diagrams (HDs) at redshifts $z = 0$--0.1 with dispersions of
only $\sim 0.15$~mag, which opened up the opportunity to measure the
expected deceleration of the Universe. At the end of the millennium,
\citet{Rie98} and \citet{Per99} reported independent observations of
high-redshift SNe ($z = 0.1$--0.5) which permitted the measurement of
the expansion history of the Universe over $\sim$~5~Gyr of lookback
time. Contrary to expectations, these observations revealed that the
Universe is currently described by an accelerated expansion. The
discovery of cosmic acceleration, which was subsequently confirmed by
additional studies of SNe~Ia \citep[e.g.,][]{Ast06,Woo07,Hic09,Fre09},
implies the possible existence of a cosmological constant, a concept
initially introduced by Albert Einstein at the beginning of the 20th
century.  Its origin is still a mystery, and there are many other
possible candidates for the repulsive ``dark energy'' that permeates
the Universe (see \citealt{Fri08} for a review).

The acceleration of the Universe has now been indirectly confirmed by
other independent experiments such as the Wilkinson Microwave
Anisotropy Probe \citep[WMAP;][]{Spe07,Ben03} and baryon acoustic
oscillations \citep[BAO;][]{BlG03,SeE03}, as well as more directly
with the integrated Sachs-Wolfe effect \citep[e.g.,][]{BoC04,Fos03}
and X-ray observations of clusters of galaxies \citep{All08}, but it
is nevertheless important to obtain additional independent
confirmation of the SN~Ia results. Although SNe~IIP are not as
luminous and uniform as SNe~Ia, \citet{HaP02} showed that the
luminosities of the former can be standardized to levels of 0.4~and
0.3~mag in the $V$ and $I$ bands, respectively. This finding converted
these objects into precise distance indicators and potentially into
useful tools to measure cosmological parameters. Moreover, SNe~IIP
would be more abundant at high redshift than SNe~Ia, since the
star-formation rate increases with redshift until $z \approx 2.5$ (see
\citealt{Hop06}, and references therein).

Type IIP supernovae provide two routes to distance determinations.
First, the ``expanding photosphere method'' \citep[EPM;][]{KiK74} is a
technique based on theoretical models that is independent of the
extragalactic distance scale \citep{Eas96,DeH05}. In EPM the spectral
energy distribution of the SN is approximated as a black body, and
atmosphere models permit one to correct the dilution of the
photospheric flux relative to a Planck function due to electron
scattering. From a detailed analysis of the properties of the SN
atmospheres, \citet{Eas96} showed that the most important variable in
determining the “dilution factors” is the effective temperature. This
result permits one to determine the photospheric angular radius of the
SN from a measurement of the color temperature, and infer the distance
from the expansion velocity measured from spectral lines, without the
need to craft specific models for each SN. This method can achieve
dispersions of 0.3~mag in the HD, which translates into a 14\%
precision in relative distances \citep{Sch94,Ha01b,Jon09}. Another
more accurate approach consists of adopting corrections and color
temperatures from tailored models for each observation, revealing an
accuracy of $\sim$~10\% in distance determination of individual SNe
\citep{DeH06,Des08}.

Second, the ``standardized candle method'' (SCM) is an empirical
technique initially proposed by \citet{HaP02} based on an
observational correlation between the absolute magnitude of the SN and
the expansion velocity of the photosphere, the
luminosity/expansion-velocity (LEV) relation. This correlation shows
that SNe~IIP with greater luminosities have higher expansion
velocities, which permits one to correct for the large ($\sim$~4~mag)
luminosity differences displayed by these objects to levels of only
$\sim 0.3$~mag. So far, the SCM has been applied to 24~low-redshift
($z<0.05$) SNe \citep{Ha03b,Ham05} and more recently by \citet{Poz09}
to 17~low-redshift ($z<0.03$) SNe, as well as by \citet{Nug06} to
5~high-redshift ($z<0.29$) SNe. The latter work was the first attempt
to derive cosmological parameters from SNe~IIP and demonstrated the
enormous potential of this class of objects as cosmological probes.

Several major sky-survey programs now in progress or to be deployed in
the upcoming years, such as the Palomar Transient Factory
\citep[PTF;][]{Rau09,Law09}, the Panoramic Survey Telescope and Rapid
Response System \citep[Pan-STARRS;][]{Hod04}, the Large Synoptic
Survey Telescope \citep[LSST;][]{Tys03}, the Visible and Infrared
Survey Telescope \citep[VISTA;][]{Eme04}, the VLT Survey Telescope
\citep[VST;][]{Cap03}, the Dark Energy Survey \citep[DES;][]{Cas07},
and the SkyMapper \citep{Gra06}, offer the promise of discovering SNe
by the thousands. Thus, we will be inevitably confronted by enormous
amounts of data on SNe~IIP which will contain potentially valuable
cosmological information. The challenge is to find an implementation
of SCM that does not require spectroscopic data since these surveys
will be purely photometric.

In spite of the great potential shown by the SCM, it still suffers
from a variety of problems which need to be addressed: (1)~the lack of
a well-defined maximum in the light curves has prevented us from
defining the phase of each event; (2)~each SN shows a different color
evolution, which has compromised the use of the photometric data for
the determination of host-galaxy extinction; (3)~the assumption for
dereddening employed by \citet{HaP02}, namely that all SNe reach the
same asymptotic color toward the end of the plateau phase, has often
led to negative extinction corrections, so it still needs to be
further tested; and (4)~the small sample size used so far, especially
the scarcity of SNe in the Hubble flow, has prevented a proper
determination of the intrinsic precision of the method.

With these obstacles in mind, we have developed a robust mathematical
procedure to fit the light curves, color curves, and velocity curves,
in order to obtain a more accurate determination of the relevant
parameters required by the SCM (magnitudes, colors, and ejecta
velocities), as described in detail in the thesis work of
\citet{Oli08}. Here we apply this technique to 37 SNe~IIP in order to
construct a HD, evaluate the accuracy of the SCM, and obtain an
independent determination of the Hubble constant. Since our sample has
several objects in common with the recent EPM analysis of
\citet{Jon09}, we perform a comparison between SCM and EPM.

We organize this paper as follows. In \S~\ref{DATA} we describe all of
our observational material including telescopes, instruments, and
surveys involved.  The analysis, methodology, and procedures, such as
the $A_{\rm G}$, $K$, and $A_{\rm host}$ corrections, are explained in
detail in \S~\ref{METH}.  The dereddening analysis, the HD, the value
of the Hubble constant, and the comparison between SCM and EPM
distances are addressed in \S~\ref{ANLS}.  The final remarks in
\S~\ref{DISC} provide a discussion about variations of the reddening
law for SNe~IIP and the possibility of a nonstandard extinction law in
the SN host galaxies. We present our conclusions in \S~\ref{CONC}.

%% file: c2Data.tex
\section{OBSERVATIONAL MATERIAL} \label{DATA}

This work makes use of data obtained in the course of four systematic
SN follow-up programs carried out during the years 1986--2003: (1)~the
Cerro Tololo SN program (1986--1996); (2)~the Cal\'an/Tololo SN
program (CT, 1990--1993); (3)~the Optical and Infrared Supernova
Survey (SOIRS, 1999--2000); and (4)~the Carnegie Type~II Supernova
Program (CATS, 2002--2003).  As a result of these efforts, photometry
and spectroscopy (some infrared but mostly optical) was obtained for
nearly 100~SNe of all types, 51~of which belong to the Type~II
class. Of these 51 SNe~II, a subset of 33 objects meets the
requirements for this study. Next we describe in general terms the
data acquisition and reduction procedures. For more details the reader
can refer to M.~Hamuy et al. (2010, in preparation).

\subsection{Photometric Data}

The photometry was acquired with telescopes from Cerro Tololo
Inter-American Observatory (CTIO), Las Campanas Observatory (LCO), the
European Southern Observatory (ESO) in La Silla, and Steward
Observatory (SO). Many different telescopes and intruments were used
to generate this dataset as shown in Table~\ref{TbIns}. In all cases
we employed CCD detectors and standard Johnson-Kron-Cousins-Hamuy
$UBVRIZ$ filters \citep{Joh66,Cou71,Ha01a}.

The images were processed with IRAF\footnote{IRAF, the Image Reduction
  and Analysis Facility, is distributed by the National Optical
  Astronomy Observatory, which is operated by the Association of
  Universities for Research in Astronomy (AURA), Inc., under
  cooperative agreement with the National Science Foundation (NSF);
  see \url{http://iraf.noao.edu}.}  through bias subtraction and
flat-fielding. All of them were further processed through the step of
galaxy subtraction using template images of the host
galaxies. Photometric sequences were established around each SN based
on observations of Landolt and Hamuy standards
\citep{Lan92,Ham92,Ham94}. The photometry of all SNe was performed
differentially with respect to the local sequence on the
galaxy-subtracted images.  The transformation of instrumental
magnitudes to the standard system was done by taking into account a
linear color term and a zero point.  Although this procedure partially
removes the instrument-to-instrument differences in the SN magnitudes,
it should be kept in mind that significant systematic discrepancies
can still remain owing to the nonstellar nature of the SN spectrum
\citep[e.g.,][]{Ham90}.

Complementary photometry for SN~2003gd obtained by \citet{vDy03} and
\citet{Hen05} was also incorporated in our analysis.

\subsection{Spectroscopic Data}

The spectroscopic data were also obtained with a great variety of
instruments and telescopes, as shown in Table~\ref{TbIns}. The
observations consisted of the SN observation immediately followed by
an emission-line comparison lamp taken at the same position in the
sky, and 2--3 flux standards per night from the list of
\citet{Ham92,Ham94}.  We always used CCD detectors, in combination
with different gratings/grisms and blocking filters. The reductions
were performed with IRAF and consisted of bias subtraction,
flat-fielding, one-dimensional (1D) spectrum extraction and sky
subtraction, wavelength calibration, and flux calibration. No attempts
were made to remove the telluric lines.

Our spectroscopic database was complemented with spectra obtained by
the CfA SN program with the FAST CCD spectrograph \citep{Fab98} at the
1.5-m telescope of the Whipple Observatory (see \citealt{Mat08} for
details on the data reduction), and by the University of California,
Berkeley SN program with the Kast double CCD spectrograph
\citep{Mil93} on the Shane 3-m telescope at Lick Observatory. The
above standard procedures, including removal of telluric absorption
lines, were employed to reduce the Lick spectra
\citep[e.g.,][]{F2003}.

\subsection{Subsample Used for this Work}\label{sample}

Of the 51~SNe~II observed in the course of these four surveys, a
subset of 33~objects comply with the requirements of (1)~being a
member of the plateau subclass; (2)~having sufficient spectroscopic
temporal coverage; and (3)~having light curves with good temporal
coverage. To this sample we added four SNe from the literature:
SN~1999gi, SN~2004dj, SN~2004et, and SN~2005cs. Table~\ref{Tb1} lists
our final sample of 37~SNe~IIP.  For each SN this table includes the
name of the host galaxy, the equatorial coordinates of the SN, the
heliocentric redshift and its source, the reddening due to our own
Galaxy \citep{SFD98}, and the survey during which the SN was observed.

%% file: c3Meth0.tex
\section{METHODOLOGY AND PROCEDURES} \label{METH}

In the thesis work of \citet{Oli08}, we describe in detail the
techniques for correcting the photometry for Galactic absorption
($A_{\rm G}$), $K$-correction terms, and host-galaxy dust absorption
($A_{\rm host}$) --- the so-called ``$AKA$ corrections''. In brief, we
employ a library of 196 spectra of SNe~II (including other SN~II types
than just SNe~IIP) to synthesize $K$-terms and absorption coefficients
given the specific redshift and $A_V$ value of each SN. Synthetic
$B-V$ and $V-I$ colors are used as proxies for the spectral energy
distribution in order to account for the evolution of each SN.

\input{c3Meth1Fits}

\input{c3Meth2HExt}

%% file: c3Meth1Fits.tex
\subsection{Fits to Light, Color, and Velocity Curves}

In the first incarnation of the SCM, \citet{HaP02} measured all of the
relevant quantities (magnitudes, colors, and velocities) at fixed
epochs with respect to the time of explosion. In most cases, however,
it proves hard to constrain this time, thus hampering the task to
compare data obtained for different SNe. It would be ideal to have a
conspicuous feature in the light curves; unfortunately, unlike other
SN types, SNe~IIP generally do not show a well-defined maximum during
the plateau phase.

One way around this is to use the end of the plateau as an estimate of
the time origin for each event. In practice it is not easy to measure
this time owing to the generally coarser sampling of the light curves
at this phase.  Thus, our first aim is to implement a robust
light-curve fitting procedure in order to obtain a reliable time
origin to be used as a uniform reference epoch to measure magnitudes,
colors, and expansion velocities. In the remainder of this section, we
proceed to implement fitting methods to measure reliable colors and
expansion velocities.

\subsubsection{$BVRI$ Light-Curve Fits}\label{LCF}

Figure~\ref{FgLCfit} shows the $BVRI$ light curves of SN~1999em. As
can be seen, there are three distinguishable phases in the light
curves.

\begin{description}

\item[---] The {\it plateau phase}, in which the SN shows an almost
  constant luminosity during the first $\sim$~100~days of its
  evolution. This phase corresponds to the optically thick period in
  which a hydrogen recombination wave recedes in mass coordinate,
  gradually releasing the internal energy of the star (see
  \citealt{KiK74,Nad03,Utr07,Ber09}, and references therein).

\item[---] The {\it linear} or {\it radioactive tail}, a linear decay
  in magnitude (exponential in flux) starting about 130~days after the
  explosion. This phase corresponds to the optically thin period
  powered by the $^{56}$Co~$\rightarrow$~$^{56}$Fe radioactive decay
  \citep[e.g.,][]{WeW80}.

\item[---] A {\it transition phase} of $\sim$~30~days between the
  plateau and linear phases.

\end{description}

Both the plateau and linear phases are trivial to model if taken
separately, but the abrupt transition makes the fitting task much more
challenging, especially with coarsely sampled light curves. After
considering several options we ended up using the arithmetic sum of
the three functions shown in Figure~\ref{FgLCfit}, as follows.

\begin{description}

\item[(1)] A Fermi-Dirac function (dotted line in
  Figure~\ref{FgLCfit}), which provides a very good description of the
  transition between the plateau and radioactive phases:
  \begin{equation}
    f_{\rm FD}(t)=\frac{-a_0}{1+e^{(t-t_{\rm PT})/w_0}},
  \end{equation}
  \noindent
  where $a_0$ represents the height of the step in units of magnitude,
  $t_{\rm PT}$ corresponds to the middle of the transition phase and
  is a natural candidate to define the origin of the time axis, and
  $w_0$ quantifies the width of the transition phase. At $t=t_{\rm
    PT}-3w_0$ the height of the step has been reduced by~4.7\%, and it
  decreases down to~95.3\% at $t=t_{\rm PT}+3w_0$.

\item[(2)] A straight line (long-dashed line in Figure~\ref{FgLCfit})
  which accounts for the slope due to the radiactive decay:
  \begin{equation}
    l(t)=p_0\, (t-t_{\rm PT}) + m_0,
  \end{equation}
  \noindent
  where $p_0$ corresponds to the slope of radioactive tail and an
  approximate slope for the plateau in units of mag per day, and $m_0$
  corresponds to the zero point in magnitude at $t=t_{\rm PT}$.

\item[(3)] A Gaussian function (short-dashed line in
  Figure~\ref{FgLCfit}), which serves mainly for fitting the $B$
  light-curve bump and the $I$ light-curve curvature during the
  plateau phase:
  \begin{equation}
    g(t)=-P\,e^{-\left(\frac{t-Q}{R}\right)^2},
  \end{equation}
  \noindent
  where $P$ is the height of the Gaussian peak in units of magnitudes,
  $Q$ is the center of the Gaussian function in days, and $R$ is the
  width of the Gaussian function.  The Gaussian function is also
  useful for reproducing the small-scale features that can appear in
  the $V$-band plateau.

\end{description}

The resulting analytic function we use to model the light curves is
the sum of the three functions detailed above,
\begin{eqnarray}
  F(t)&=&f_{\rm FD}(t)+l(t)+g(t)\nonumber\\
  &=&\frac{-a_0}{1+e^{(t-t_{\rm PT})/w_0}}+p_0\,
  (t-t_{\rm PT})+m_0-P\,e^{-\left(\frac{t-Q}{R}\right)^2},
\label{fnc_eq}
\end{eqnarray}
\noindent
which has eight free parameters. This function is fitted to the
individual light curves in unison (not by parts) using a $\chi^2$
minimizing procedure. In order to find the minimum $\chi^2$, we use
the Downhill Simplex Method \citep{NeM65}.  Despite being not very
efficient in terms of the number of iterations required, it provides
robust solutions. Examples of the analytic fits are shown as solid
lines in Figure~\ref{FgLCfit}.  Although we cannot model most of the
small-scale features in the plateau, the fitting does a very good job
modelling the transition. Furthermore, the analytic function gives us
important parameters that characterize the light-curve shape,
particularly $t_{\rm PT}$ which provides a time origin. In the plots
in Figure~\ref{FgLCfit}, the time axis is chosen to coincide with the
value of this parameter obtained from the $V$~light curve. In
Table~\ref{TbAnPar} we gather the eight parameters and the
corresponding reduced $\chi^2$ ($\chi_{\nu}^2$) of the $V$~light curve
fitting for the whole SN sample.

A critical quantity in the analysis that follows is $t_{\rm PT}$ and
its uncertainty, both of which determine the uncertainties in all of
the relevant SCM quantities. For most cases, when the tail phase has
been observed, the $\chi^2$ minimizing routine has no difficulties
finding $t_{\rm PT}$ and delivers a credible error.  On the other
hand, when the light curve does not have any late-time data points,
the routine underestimates $\sigma(t_{\rm PT})$, in which case we need
to provide a more realistic estimate of this uncertainty. Only for a
few cases did we have to arbitrarily set $t_{\rm PT}$ and its error
based on the light-curve sampling (see Column~3 in
Table~\ref{TbAnPar}).

Regardless of the sampling of the light curves, we assign a minimum of
$\sigma(t_{\rm PT})=2$~days. Although these uncertainties in $t_{\rm
  PT}$ are very conservative and arbitrary, they do not translate into
high or unrealistic errors in quantities such as magnitudes, colors,
or expansion velocities, as we will see later.

\subsubsection{Color-Curve Fits}\label{CC}

Figure~\ref{FgColr} shows the $B-V$, $V-R$, and $V-I$ colors of three
prototypical SNe~IIP corrected for $A_{\rm G}$ and $K$-terms as
described by \citet[Chapter~3.1]{Oli08}. The time origin in the
abscissa corresponds to $t_{\rm PT}$, the middle of the transition
phase. In each case we employ all the data points between days $-100$
and~$-10$ to fit a Legendre polynomial shown with solid lines in
Figure~\ref{FgColr}. The degree of the polynomial was chosen on a
case-by-case basis and varied between third and sixth order. It is
evident that, during the plateau phase, the photosphere gets redder
with time owing to the decrease of the surface temperature as the SN
expands and to a redistribution of flux from blue to red wavelengths
due to line blanketing. According to the theory, the photospheric
temperature should approach and never get below the temperature of
hydrogen recombination around 5000~K. Based on this physical argument,
\citet{HaP02} argued that all SNe~IIP should reach the same intrinsic
colors toward the end of the plateau phase; hence, they proposed that
the color excesses measured at this phase could be attributed to dust
reddening in the SN host galaxy and be exploited to measure $A_{\rm
  host}$. \citet{HaP02} performed their analysis with a simple
naked-eye estimate of the asymptotic colors. Here we improve
significantly this situation through the formal color-curve fitting
procedure described above.

Armed with the polynomial fits, we proceeded to interpolate colors on
a continuous grid with one-day spacing between days $-80$ and~$-10$
for analyzing colors at multiple epochs (see \S~\ref{aho}). In a
handful of cases, the data did not encompass the whole grid and we had
to extrapolate colors, but never by more than 3~days from the nearest
data point.

\subsubsection{\ion{Fe}{2}~Based Expansion-Velocity Curves}\label{exp}

The third ingredient for SCM is the velocity of the SN ejecta.
Different spectral lines yield different expansion velocities
depending on the region from which the photon is emitted. The
\ion{Fe}{2}~$\lambda5169$ line is thought to be a good proxy for
SN~IIP photospheric velocity \citep{DeH05} and has been previously
employed for SCM. Here we use the \ion{Fe}{2} expansion velocities
measured by \citet{Jon09} from the minimum of the P-Cygni line profile
after correcting the spectra for the heliocentric redshifts of the
host galaxies. The uncertainties of these measurements were estimated
to be $\sim 85$~\kmpsec\ from the quality of the wavelength
calibration and the signal-to-noise ratio (S/N) of the absorption
feature of the P-Cygni line profile. Although these measurements do
not consider relativistic corrections, the latter prove negligible
($\lesssim 40$~\kmpsec) in the range considered here
(2000--5000~\kmpsec).

Figure~\ref{FgVels} shows \ion{Fe}{2} velocities as a function of SN
phase, for four different SNe selected for their wide range of
sampling characteristics. As shown by \citet{HaP02}, the
\ion{Fe}{2}~$\lambda5169$ expansion velocity curve during the plateau
phase can be properly modeled with a power law, which we choose to be
of the form

\begin{equation}
  \upsilon_{\rm FeII}(t) = A \, (t-t_0)^\alpha,
\label{vel_eq}
\end{equation}

\noindent
where $A$, $t_0$, and $\alpha$ are free fitting parameters without
obvious physical meaning.

In general, we fit for all three parameters, but when only two
velocity measurements are available (e.g., SN~1992af in the top-left
panel of Figure~\ref{FgVels}) we fix the $\alpha$ exponent to~$-0.5$,
which corresponds to a typical value for our sample. As shown with
solid lines in Figure~\ref{FgVels}, the fits are quite
satisfactory. Here we choose to restrict the power-law fits to the
plateau phase, since the power-law behavior is not observed for
expansion velocities beyond the transition phase.  We use the same
one-day continuous grid as for the color curves in order to
interpolate velocities at different epochs between $-80$ and
$-10$~days.  Given the good quality of the fits and the shallow slope
at late epochs, we allow extrapolations of up to 15~days past the
nearest data point (see SN~1999cr in Figure~\ref{FgVels}). At the
early-time boundary, we reduce the extrapolations to 10~days prior to
the first point, because the power law becomes steeper at early times
(e.g., see SN~2003gd in Figure~\ref{FgVels}). When only two velocity
measurements are available we reduce the extrapolations by
5~days. Since at the end of the plateau phase the SN atmosphere starts
to become nebular, we discarded data points beyond the transition in
the velocity fits. We do not impose any constraints on the first data
point.

%% file: c3Meth2HExt.tex
\subsection{Host-Galaxy Extinction Determination}\label{aho}

To determine the host-galaxy extinction, we assume that all SNe~IIP
should evolve from a hot initial stage to one of uniform photospheric
temperature, owing to the hydrogen-recombination nature of their
photospheres.

Although simple in theory, there are a few practical difficulties when
comparing the color evolution. First, it is not always possible to
constrain the time of explosion and accurately compare color curves of
different objects.  One way around this is to use the transition time,
$t_{\rm PT}$, defined in \S~\ref{LCF}. The second problem is
illustrated in Figure~\ref{FgColr}: the color-curve shapes can vary
significantly from one SN to another, preventing one from measuring a
single color offset between two SNe. To circumvent this, we pick a
single fiducial epoch in the color evolution and assess the
performance of the color as a reddening estimator. Our polynomial fits
to the color curves are very convenient for this purpose, since they
allow us to interpolate reliable colors on a day-to-day basis over a
wide range of epochs and explore which epoch gives the best results.

If all SNe share the same intrinsic temperature at some epoch, we
expect the subset of dereddened SNe to have nearly identical colors
($C_0$); the remaining objects should show color excesses,
$E(C)=C-C_0$, in direct proportion to their extinctions. A useful
diagnostic to check our underlying assumption is the color-color plot:
unreddened SNe should occupy a small region in this plane. If we
further assume the same extinction law in the SN host galaxies, the
subset of extinguished objects should describe a straight line
originating from such a region.  The figures of merit in this test are
(1)~the color dispersion displayed by the unreddened SNe, (2)~the
slope described by the reddened SNe (which is determined by the
extinction law), and (3)~the dispersion relative to the straight line
(the smaller the better).

We have identified four objects in our sample (SN~2003B, SN~2003bl,
SN~2003bn, SN~2003cn) consistent with zero reddening. These objects
were selected for having (1)~no significant \nad\ interstellar lines
in their spectra at the redshifts of their host galaxies, and
(2)~early-time spectra that show no signs of host-galaxy extinction.
To judge the latter we used extinction values determined from fits of
SN~IIP atmosphere models to our early-time spectra as done by
\citet{DeH06} and \citet{Des08}. Such models use the SN spectral lines
to constrain the photospheric temperature and the continuum to
restrict the amount of extinction.  Examples of fits are shown in
Figure~\ref{FgSpec}.

We investigated two color-color plots ($V-I$ versus $V-R$, and $V-I$
versus $B-V$) over a wide range of epochs (from day $-50$ to~$-15$)
{\it after correcting the photometry for Galactic extinction and
  $K$-terms} as described by \citet{Oli08}. The best results were
found for the $V-I$ versus $V-R$ diagram constructed from day~$-30$
and shown in Figure~\ref{FgVIVR}. Fortunately, all SNe in our sample
have enough data to interpolate color curves at day~$–30$. At this
epoch, approximately the end of the plateau, we obtain the linear
behavior expected for a sample with the same intrinsic color but
different degrees of extinction. Shown with a triangle is the single
SN consistent with zero extinction which is, remarkably, one of the
bluest objects in this diagram; the other three unextinguished SNe do
not have $R$-band photometry. A least-squares fit to the data yields a
slope of $1.59\pm0.09$, which is lower than the $E(V-I)/E(V-R)=2.12$
ratio expected for a Galactic extinction law ($R_V=3.1$) shown as a
vector in Figure~\ref{FgVIVR}.  This suggests a somewhat different
extinction law in the SN hosts compared to the Galaxy.  The dispersion
of 0.059 mag in $V-I$ is a promising result as it translates into an
uncertainty of $A_{\rm host}(V)\,=\,0.15$~mag, which corresponds to
the limiting precision of this method. The reduced $\chi^2$ of 1.55
implies that the dispersion can be accounted almost solely by our
error bars and that any instrinsic color dispersion in our sample is
$\leq 0.06$ mag. The bottom line is that both the $V-R$ and $V-I$
colors fulfill the minimum requirements as reddening indicators.

The best results from the $B-V$ versus $V-I$ analysis, obtained from
day~$-30$, are shown in Figure~\ref{FgVIBV}. The four SNe consistent
with zero extinction, plotted with triangles, have weighted average
colors of $(B-V)_0=1.147\pm0.053$ mag and $(V-I)_0=0.656\pm0.053$ mag,
where the dispersion was used as measure of the uncertainty.  Note
that there are five SNe in this diagram which are slightly bluer than
the unreddened sample. The $V-I$ color dispersion of 0.076 mag is
greater than that obtained in Figure~\ref{FgVIVR} and is most likely
due to the $B-V$ color, since the $B$ band is more sensitive to the
metallicity of the SN, owing to many absorption lines that lie in this
spectral region. Therefore, we believe that the greater dispersion in
this diagram could be due to the different metallicities of our SN
sample. A least-squares fit to the data yields a slope of
$0.77\pm0.04$. This slope is quite different from the
$E(V-I)/E(B-V)=1.38$ ratio expected for the Galactic extinction law
(shown as a vector in Figure~\ref{FgVIBV}), in agreement with the
departure from a standard reddening law seen in the $V-I$ versus $V-R$
diagram.

We conclude from our exploration that, while the $B-V$ color is
problematic, both the $V-R$ and $V-I$ colors offer a promising route
for dereddening purposes. In what follows, we will employ solely the
$V-I$ color since only a small subset of our objects possess $R$-band
photometry. Although the evidence points to a non-Galactic reddening
law, for now we will assume a standard reddening law (later we will
relax this assumption; see \S~\ref{HDRV}).

Using our library of SN~II spectra, we computed synthetically the
appropriate conversion factor between $E(V-I)$ and $A_V$ for SNe~II
and a standard reddening law ($R_V=3.1$), yielding

\begin{equation}
  \beta_V=\frac{A_V}{E(V-I)}=2.518.
\label{beta_eq}
\end{equation}

\noindent
The host-galaxy extinction can be computed from
$A_V(V-I)=\beta_V\,[(V-I)-(V-I)_0]$ and, assuming an intrinsic color
of $(V-I)_0=0.656\pm0.053$ mag, its numerical expression with the
corresponding uncertainty is

\begin{eqnarray}
  A_V(V-I)&=&2.518\,[(V-I)-0.656],\label{Ext_eq} \\
  \sigma(A_V)&=&2.518\,\sqrt{\sigma_{(V-I)}+0.053^2+0.059^2},\nonumber
\end{eqnarray}

\noindent
where $V-I$ corresponds to the color of a given SN at day~$-30$
(corrected for $K$-terms and foreground extinction) and
$\sigma_{(V-I)}$ combines the instrumental errors in the $V$ and $I$
magnitudes, the uncertainty in $A_{\rm G}$ and the $K$-terms
\citep[see][Chapters~3.1.1 and 3.1.2, respectively]{Oli08}, and the
uncertainty in $t_{\rm PT}$ (\S~\ref{LCF}). The uncertainty in the
intrinsic $(V-I)_0$ color was also included in the error of $A_V$
along with the $V-I$ standard deviation in the $V-I$ versus $V-R$
diagram.

The host-galaxy reddening values obtained from this technique are
listed in Column~5 of Table~\ref{Tb2} (with the uncertainties given in
parentheses for the whole sample of 37~SNe). There are eight SNe with
$V-I$ colors bluer than $(V-I)_0$ which stand out in this table for
their inferred negative reddenings. Although not physically
meaningful, these negative values are statistically consistent with
zero or moderate reddenings. In fact, seven of these eight objects
differ by $<1.1\sigma$ from zero reddening. Only SN~1992af differs by
$1.8\sigma$ from $A_{\rm host}=0$ mag.

As a comparison, \citet{Kri09} employed a similar procedure to
determine the extinction to SN~2003hn. While we get $A_V=0.46\pm0.21$,
they obtained $A_V=0.54\pm 0.14$, values that are consistent within
the uncertainties. In the following section we compare this method
with other dereddening techniques.

%% file: c4Anly0.tex
\section{ANALYSIS} \label{ANLS}

\input{c4Anly1DeTc}
\input{c4Anly2LmVe}
\input{c4Anly3HDgr}
\input{c4Anly4HCst}
\input{c4Anly5Dist}

%% file: c4Anly1DeTc.tex
\subsection{Comparing Dereddening Techniques} \label{dtc}

SN~IIP atmosphere models were fitted to our library of spectra as done
by \citet{DeH06} and \citet{Des08}. In these fits the spectral lines
are used to constrain the photospheric temperature and the
corresponding continuum is employed to estimate the extinction. Thus,
the quality of the fit depends on (1) the determination of the
photospheric temperature by means of the spectral lines, and (2) how
well the continuum is represented by the extinguished blackbody
emission. A crucial condition for this technique to work is the
spectrophotometric quality of the spectra.

In general, our observations were obtained with the slit oriented
along the parallactic angle \citep{Fil82} and the relative shape of
the spectra should be accurate. However, this was not always possible;
moreover, sometimes the spectra were contaminated by light from the
host galaxy. Accordingly, we first checked the flux calibration of
each spectrum by synthesizing magnitudes and comparing them to the
observed magnitudes, duly interpolated to the time of the
spectroscopy. We typically found an agreement better than 0.05~mag
between the observed and synthetic colors, thus confirming our
confidence in the flux calibration. In the 64~cases (18\% of all
spectra) where we found significant differences between synthetic and
observed colors ($\geq$~0.1 mag for any color), we applied a low-order
polynomial correction to the spectrum. Basically, this ``mangling''
correction used the observed photometry to change the slope of a
spectrum. After checking the flux calibration (and mangling it, if
needed), the next step was to correct for Galactic absorption and
deredshift the spectra. This database was then used for the atmosphere
model fits. Examples of the fits are shown in Figure~\ref{FgSpec}.

The resulting spectroscopic reddenings are summarized in Column~2 of
Table~\ref{Tb2}. As pointed out by \citet{DeH06}, the spectrum-fitting
technique works much better with early-time spectra than with
late-time observations. At late times the photosphere has receded in
mass, exposing inner and more metal-rich layers, which translates into
an overabundance of heavy-element absorption lines. Therefore, the
fitting to the continuum (essentially a temperature fitting) is
hampered by the increased line opacity at late times. We thus divided
our sample into four subcategories based on the epoch and the flux
quality of the spectra used in the reddening determination: (a) {\it
  gold:} early-time spectra, without mangling correction; (b) {\it
  silver:} early-time spectra, with mangling correction; (c) {\it
  bronze:} late-time spectra, without mangling correction; and (d)
{\it coal:} late-time spectra, with mangling correction. Note that the
main criterion is whether the spectrum was obtained at early or late
times, and the second criterion corresponds to the flux-calibration
quality.  We consider the unmangled spectra to be of higher quality
than the mangled ones, because the former do not require any
corrections and suggest that the observations were better performed.

The above subclassification is given for each SN in Column~3 of
Table~\ref{Tb2}. Note that the uncertainty is directly related to this
subclassification. \citet{Des08} estimate an error of
$\sigma_{E(B-V)}=0.05$~mag when using early-time spectra, and
$\sigma_{E(B-V)}=0.10$~mag when using late-time spectra. We refine
this argument, ramping up from 0.05 to 0.10 mag, depending on the
number of spectra employed for each extinction determination.  These
values are given in Table~\ref{Tb2} in units of $A_V=3.1\, E(B-V)$.

Figure~\ref{FgRd1} shows a comparison between the spectroscopic
reddenings $A_V$(spec) and our color reddenings $A_V(V-I)$ derived in
\S~\ref{aho}, for the 17~SNe belonging to the top three subclasses
({\it gold} + {\it silver} + {\it bronze}). Good agreement is found
between both techniques, with a dispersion of 0.38~mag. The resulting
$\chi_{\nu}^2=1.4$ suggests that this dispersion is consistent with
the combined errors of both techniques. The exceptions are two
objects, SN~1999br and SN~2003ho.  SN~1999br has only plateau-phase
photometry, making it hard to determine $t_{\rm PT}$. Although the
uncertainty in $t_{\rm PT}$ is quite large (20~days), this does not
have a great impact on the error of the $V-I$ color due to the
flatness of the color curve at these epochs (+0.0024 mag per
day). Another possible reason for the disagreement could be an
intrinsically redder $V-I$ color than that of the bulk of the SNe,
caused by the extreme properties (low luminosity and expansion
velocity) of SN~1999br.  The second discrepant object (SN~2003ho)
belongs to the {\it bronze} group, so it is possible that the
difference could be due to the use of a late-time spectrum in the
determination of the spectroscopic reddening.

Another way to estimate host-galaxy reddening is from the
\nad\ $\lambda\lambda$5889, 5896 interstellar absorption doublet
observed in the SN spectrum at the host-galaxy redshift.  Whenever the
line was detected we measured its equivalent width; in those cases
where we did not detect the \nad\ line we assigned it a null value. In
all cases we estimated the uncertainty in the equivalent width (EW)
based on the S/N of the continuum around this line. We converted these
measurements into visual extinctions, $A_V$(\nad), using the
calibration of \citet{Bar90}:

\begin{equation}
  E(B-V)\,\approx\,0.25\, {\rm EW}(\mbox{\nad}),\label{EW2EBV}
\end{equation}

\noindent
where the units of $E(B-V)$ and EW are mag and \AA, respectively.
Although \citet{Tur03} showed that this relation is bimodal, and
\citet{Blo09} found that the $E(B-V)$ estimates based on EW(\nad) were
highly inaccurate, the $E(B-V)$ values derived from
Equation~\ref{EW2EBV} (tabulated in Column~4 of Table~\ref{Tb2})
provide a rough estimate of $E(B-V)$ to compare with our results. The
comparison between $A_V$(\nad) and our technique, shown in
Figure~\ref{FgNad}, exhibits a dispersion of 0.53~mag
($\chi_{\nu}^2=3.8$), which is much higher than the $\sigma =
0.38$~mag obtained from the previous comparison. The disagreement can
be attributed to the fact that the absorption line traces the gas
content along the line of sight, but does not necessarily probe dust
\citep{MuZ97}. Furthermore, reddenings derived from the EW of the
\nad\ lines measured from our low-dispersion spectra simply cannot be
expected to be accurate; the D lines produced by a typical
interstellar cloud are saturated \citep{Hob74}. The only way that one
can hope to derive a reliable reddening from the D lines is via very
high-dispersion spectra that resolve the lines and allow the column
density to be determined.

%% file: c4Anly2LmVe.tex
\subsection{Luminosity versus Expansion Velocity}

After developing the dereddening method based on late-time colors, we
can now revisit the LEV relation originally discovered by
\citet{HaP02}, which is at the core of the SCM. For this purpose we
applied $AKA$ corrections to our photometry
\citep[][Chapter~3.1]{Oli08}, we used our analytic fits (\S~\ref{LCF})
to interpolate $BVI$ magnitudes on day~$-30$, and we employed the
cosmic microwave background (CMB) redshifts\footnote{The CMB redshifts
  were computed by adding the heliocentric redshifts given in
  Table~\ref{Tb1} and the projection of the velocity of the Sun
  relative to the CMB in the direction of the SN host galaxy. For the
  latter we adopted a velocity of 371~\kmpsec\ in the direction
  $(l,b)=(264.1^\circ, 48.3^\circ)$ given by \citet{Fix96} plus the
  uncertainty of the Local Group velocity (187~\kmpsec).} in
Table~\ref{TbHDs} to convert apparent magnitudes to absolute values
(assuming $H_0=70$~\dimho\ from our final result). In order to assess
the dereddening technique, we preserved the relative reddening
estimations by keeping the negative values of host-galaxy
absorption. We performed a power-law fit to interpolate a velocity
contemporaneous with the photometry (day~$-30$) as described in
\S~\ref{LCF}. Only for SN~1999cr did we have to use an extrapolated
velocity (see top-right panel in Figure~\ref{FgVels}).  From our
original sample of 37~SNe~IIP we were able to use 30~SNe to build this
relation, since five of them do not have measured \ion{Fe}{2}
velocities at day~$-30$, and two others have extremely low redshifts
($cz_{\rm CMB}<300$~\kmpsec).

Figure~\ref{FgLmVe} shows the resulting LEV relation (absolute
magnitude versus expansion velocity) for all $BVI$ bands. Evidently we
recover the result of \citet{HaP02}, namely that the most luminous SNe
have greater expansion velocities.  In their case the data were
modeled with a linear function. Our sample suggests that the relation
may be quadratic, but we need more SNe at low expansion velocities to
confirm this suspicion.  Linear least-squares fits to our $BVI$ data
yield the following solutions:

\begin{eqnarray}
  M(B)_{-30}-5\,{\rm log}{(H_0/70)}&=&3.50(\pm0.30)~\log{\Bigr[\upsilon_{\rm
  FeII}(-30)/5000\Bigr]} - 16.01(\pm0.20),\\
  M(V)_{-30}-5\,{\rm log}{(H_0/70)}&=&3.08(\pm0.25)~\log{\Bigr[\upsilon_{\rm
  FeII}(-30)/5000\Bigr]} - 17.06(\pm0.14),\\
  M(I)_{-30}-5\,{\rm log}{(H_0/70)}&=&2.62(\pm0.21)~\log{\Bigr[\upsilon_{\rm
  FeII}(-30)/5000\Bigr]} - 17.61(\pm0.10),
\end{eqnarray}

\noindent
which are shown with solid lines in Figure~\ref{FgLmVe}. The relation
found by \citet{HaP02} for the $V$ band is indicated with the dashed
line in the middle panel of the same figure. Given that the study of
\citet{HaP02} was performed using data at day~50 after explosion
(around day~$-60$ on our time scale), it is not unexpected that their
LEV relation is shifted to higher expansion velocities. Some of the
difference in slope is explained by the inclusion of SN~2003bl in our
sample, which flattens the correlation. This relation exhibits a
dispersion of 0.3~mag, similar to that reported before. Such low
scatter is very encouraging as it implies that the expansion
velocities can be used to predict the SN luminosities, to standardize
them, and to derive distances.

Recently, \citet{KaW09} have computed hydrodynamical models for
SNe~IIP, which support the LEV relation initially reported by
\citet{HaP02}. Also, M.~Bersten (2010, in preparation) has used her
own hydrodynamical models to investigate the LEV relationship. By
varying the explosion energy, she has obtained an LEV relation with a
slope consistent with that reported here.

%% file: c4Anly3HDgr.tex
\subsection{Hubble Diagrams}\label{HDgr}

In order to assess the performance of the SCM, we employ (a)
host-galaxy redshifts in the CMB frame with an uncertainty of
187~\kmpsec, (b) $BVI$ apparent magnitudes and $V-I$ colors {\it
  corrected for $K$-terms and Galactic reddening}, and (c) \ion{Fe}{2}
velocities (in units of \kmpsec).  Table~\ref{TbHDs} lists these
values for the 37~SNe of our sample, of which 35 meet an initial
requirement of $cz_{\rm CMB}>300$~\kmpsec.  The weighted dispersion of
the data points around a linear fit in the HDs allows us to evaluate
the precision of different approaches.

\subsubsection{Using $A_V(V-I)$ and $A_V$(spec)}\label{HDAcs}

The top-left panel of Figure~\ref{HDB} shows a HD constructed from
$B$-band magnitudes interpolated to day~$-30$. The top-right panel
shows magnitudes additionally corrected for the LEV relation, and the
bottom-left panel includes further corrections for host-galaxy
extinction (using the $V-I$ color calibration given in
\S~\ref{aho}). In each case we perform a weighted linear least-squares
fit of the form

\begin{equation}
 m~+~\alpha\,\log{(\upsilon_{\rm FeII}/5000)}~-~A_{\rm host}
 ~=~5\,\log{cz}~+~zp,
\label{HDeq1}
\end{equation}

\noindent
where $m$ is the apparent magnitude corrected for $K$-terms and
Galactic absorption, $\upsilon_{\rm FeII}$ is the expansion velocity,
$A_{\rm host}$ is the host-galaxy absorption, and $z$ is the CMB
redshift. Note that the SN redshift enters logarithmically in this
model, implying that low-redshift SNe carry less weight in the
fit. The only fitting parameters are $\alpha$ and the zero point
($zp$), but in the top-left panel we set $\alpha=0$ and we fit only
for the $zp$. The significant decrease in the dispersion, from 0.72 to
0.36~mag, clearly demonstrates the need to add the velocity and
host-galaxy extinction terms.

An inspection of the $V$-band diagram (Figure~\ref{HDV}) shows a large
scatter of 0.54~mag in the top-left panel. When we correct for
expansion velocities, the scatter drops to 0.50~mag. This is certainly
not unexpected given the LEV relation reported in the previous
section. It is encouraging to notice that the dispersion drops from
0.50 to 0.45~mag when we include our host-galaxy extinction
corrections; this indicates the utility of our color-based dereddening
technique. The reduced $\chi^2$ value of 2.45 implies that most of the
scatter is caused by the observational errors.  We performed the same
analysis using other epochs and found that day~$-30$ yielded the
lowest dispersion.

If we turn our attention to the $I$~band (Figure~\ref{HDI}), the final
dispersion is 0.45~mag, identical to that found in the $V$ band. It
seems that the dispersion could have some dependence on wavelength,
since it decreases from 0.45 in $VI$ to 0.36~mag in $B$.  The scatter
of $\sim$~0.4 mag in $BVI$ is comparable to, but somewhat larger than,
the $\sim$~0.34 mag dispersion obtained in previous SCM studies
\citep{HaP02,Ha03b,Ham05}. It is important to notice that the
dispersions are independent of the intrinsic $(V-I)_0$ color
calculated in \S~\ref{aho}.

In the bottom-right panels of Figures~\ref{HDB}, \ref{HDV},
and~\ref{HDI} we examine the SCM using the spectroscopic reddenings
determined for a set of 28~SNe~IIP from Table~\ref{Tb2} instead of our
color-based extinctions. The resulting dispersions in $BVI$ are (0.67,
0.54, 0.42 mag), compared with (0.36, 0.45, 0.45 mag) when using
color-based extinctions. All of the fitting parameters derived from
both dereddening techniques are compiled in Table~\ref{TbPar}.

\subsubsection{Leaving $R_V$ as a Free Parameter}\label{HDRV}

The previous analysis suggests that the dispersions are somewhat
larger than the observational errors. One possible source of scatter
could be the reddening law. As shown in \S~\ref{aho} and
Figure~\ref{FgVIVR}, we have evidence for a somewhat different
extinction law in the SN hosts compared to the Galaxy.  Here we take
this idea a step further and analyze the HD leaving $R_V$ as a free
parameter. To accomplish this we model the data with the expression

\begin{equation}
  m + \alpha\,\log{(\upsilon_{\rm FeII}/5000)} - \beta
  (V-I)~=~5\,\log{cz} + zp.
\label{HDeq2}
\end{equation}

\noindent
Note that we are replacing $A_{\rm host}$ in Equation~(\ref{HDeq1})
with the term $\beta(V-I)$.  The $\beta$ factor is a new free
parameter to be marginalized along with $\alpha$ and $zp$, and $V-I$
is the color on day~$-30$ {\it uncorrected} for host-galaxy dust
absorption. This is the same approach adopted by \citet{Tri98} for
SNe~Ia but using decline rate instead of expansion photospheric
velocity. The underlying assumption in our case is that the absolute
magnitudes of SNe~IIP are a two-family function of \ion{Fe}{2}
expansion velocity and $V-I$ color. The latter is interpreted as a
color excess which varies according to dust in the host galaxy. Once
$\beta$ is known we can use it to solve for host-galaxy extinction
from $A(\lambda)=\beta\,\Bigl[\,(V-I)-0.656\,\Bigr]$, where 0.656 is
the intrinsic $V-I$ color of SNe~IIP (see Equations~(\ref{beta_eq})
and (\ref{Ext_eq}) in \S~\ref{aho}).

Figure~\ref{HDBVI} shows the $BVI$ HDs for our set of 30~SNe~IIP. We
obtain dispersions of (0.28,~0.31,~0.32 mag) in $BVI$, respectively,
compared with (0.36,~0.45,~0.45 mag) when $R_V$ is kept fixed. The
$\chi_{\nu}^2$ values increase from (1.6,~2.5,~2.5) to
(2.8,~5.4,~8.7). This change is due to the fact that we are not using
the intrinsic color, which gives the major contribution to the errors,
even more in the $I$ band. Although we expect a reduction in the
scatter due to the inclusion of an additional parameter, the large
drop in the dispersion is remarkable.  The last three lines of
Table~\ref{TbPar} show the parameters we obtain by minimizing the
dispersion in the HD for each band. When we restrict the sample to
objects with $cz_{\rm CMB}>3000$~\kmpsec\ (leaving aside SNe with
potentially greater peculiar velocities), we end up with 19~SNe in the
$B$ band and 20~SNe in the $VI$ bands. The resulting HDs in $BVI$ show
dispersions of (0.25,~0.28,~0.30 mag), respectively, and the zero
points of the HDs change by only 0.02~mag. Not surprisingly, these are
lower than the (0.28,~0.31,~0.32 mag) dispersions obtained from the
whole sample.

Efforts to remove the peculiar velocities of low-redshift galaxies
were made by means of the parametric model for peculiar flows of
\citet{Ton00}. The ingredients of this model are (1) the observed
recession velocity of the galaxy, (2) the Hubble constant, and (3) the
surface-brightness-fluctuations distance from \citet{Ton01}. The
output is an estimate of the peculiar velocity of the galaxy
\citep[see][\S~1.4 for details]{Ha03a}. After correcting 10 host
galaxies in our sample, the results are not conclusive. The scatter in
the HDs decreased only 0.01~mag for the $I$ band.

Our assumption is that the term $\beta(V-I)$ in Equation~(\ref{HDeq2})
corresponds to the extinction in a broad-band magnitude (with central
wavelength $\lambda$).  Thus, $\beta$ is the ratio $A(\lambda)/E(V-I)$
(see Equation~(\ref{beta_eq}) in \S~\ref{aho}) and is related to the
shape of the extinction law. For each of the $BVI$ bands, we used our
library of SN~II spectra to compute synthetically the value of
$\beta_\lambda$ as a function of $R_V=A_V/E(B-V)$; see
Figure~\ref{FgRv}. This allowed us to convert the $\beta_\lambda$
values resulting from our fits into $R_V$ values. Our fits yield
$\beta_\lambda=(2.67\pm0.13, 1.67\pm0.10, 0.60\pm0.09)$ for the $BVI$
bands, respectively, which translate into $R_V=(1.38^{+0.27}_{-0.24},
1.44^{+0.13}_{-0.14}, 1.36^{+0.11}_{-0.12})$.  These values are
remarkably consistent with each other and significantly lower than the
$R_V=3.1$ value of the standard reddening law.  Independent evidence
for a low-$R_V$ law has already been reported from studies of SNe~Ia
(see \S~\ref{DISC}).

The dispersion of 0.28--0.32~mag in the HDs translates into a
precision of 13--14\% in the determination of relative
distances. However, as noticed after inspection of Figure~\ref{Mcor},
the magnitude residuals are highly correlated among $BVI$ and
generally decrease with increasing redshift. The likely explanation is
that we are seeing the effects of peculiar velocities of the host
galaxies, in which case the precision of the SCM could be better than
the 13--14\% derived from the scatter in the HDs. To examine this
point, for each SN we took the $cz_{\rm CMB}$ values from
Table~\ref{TbHDs}, and calculated the distance modulus residual

\begin{equation}
  \Delta\mu = m + \alpha\,\log{(\upsilon_{\rm FeII}/5000)} - \beta
  (V-I) - 5\,\log{cz} - zp,
\label{HDres}
\end{equation}

\noindent 
for all $BVI$ bands. Figure~\ref{resCor} shows strong correlations of
these magnitude differences against each other. The root-mean square
(rms) scatter in the $\Delta\mu(B)$ vs. $\Delta\mu(V)$ relation is
0.12~mag, and for $\Delta\mu(B)$ vs. $\Delta\mu(I)$ it is
0.19~mag. This translates to 6\% and 9\% in distance, respectively. If
this correlation of the residuals is due to peculiar velocities, then
the true distance precision of SCM is 6--9\%, and not 13--14\% as
derived from the Hubble diagrams (which include scatter produced by
the peculiar velocities). This result is very close to the
$\sim$~7--10\% precision of methods using SNe~Ia
\citep[e.g.,][]{Phi93,Ha96a,RPK96,Phi99} when they were first used to
reveal the acceleration of the Universe, which demonstrates that
SNe~IIP have great potential for determining extragalactic distances
and therefore cosmological parameters.

%% file: c4Anly4HCst.tex
\subsection{The Hubble Constant}\label{H0}

The Hubble constant is a parameter of central importance in cosmology
which can be determined from our HDs. This can be accomplished as long
as we can convert apparent magnitudes into distances. The calibration
was done with two objects for which we found Cepheid distances in the
literature: SN~1999em \citep[$\mu=30.34\pm0.19$ mag;][]{Le02a} and
SN~2004dj \citep[$\mu=27.48\pm0.24$ mag;][]{Fre01}. For each
calibrating SN we can solve for $H_0$ using the expression

\begin{equation}
  \log\,H_0~=~0.2~\Bigl[\,m+\alpha\,\log{(\upsilon_{\rm
      FeII}/5000)}-\beta(V-I)-\mu-zp+25\,\Bigr]_{t=-30},
\end{equation}

\noindent
where $m$ is the apparent magnitude, $\upsilon_{\rm FeII}$ is the
expansion velocity, and $V-I$ is the color of the calibrating SN, all
measured on day~$-30$ and given in Table~\ref{TbHDs}; $\alpha$,
$\beta$, and $zp$ are the fitting parameters given in the bottom three
lines of Table~\ref{TbPar}.

Table~\ref{TbHo} summarizes our calculations for the two calibrators.
We see that the corrected $BVI$ absolute magnitudes of SN~1999em and
SN~2004dj differ from each other by 0.86--1.13~mag.  This is
considerably greater than the scatter of 0.3~mag in the LEV relation,
but statistically not impossible; see Figure~\ref{Mcor}, where we plot
with triangles the corrected absolute magnitudes of the two
calibrating SNe, on top of the whole sample of SNe employed in the
HDs. This difference leads to $H_0$ values in the range
62--105~\dimho. With only two calibrating SNe, the SCM still has much
room for delivering a more accurate and precise value for $H_0$.

%% file: c4Anly5Dist.tex
\subsection{Distances}\label{DIST}

We are in a position to calculate distances to all of the SNe of our
sample. This can be done with the expression

\begin{equation}
  \mu=m+\alpha\,\log{(\upsilon_{\rm FeII}/5000)}-\beta
  (V-I)-\langle M_{\rm corr}\rangle,
  \label{muEq}
\end{equation}

\noindent
where $\langle M_{\rm corr}\rangle$ is the corrected absolute
magnitude of SNe~IIP on day~$-30$. For this purpose we employ the
weighted mean of the two calibrating SNe given in
Table~\ref{TbHo}. The resulting values are given in Table~\ref{TbMu}
for 32~SNe~IIP; the last column shows the weighted average of their
distance moduli and its uncertainty, assuming no covariance.  As
Equation~\ref{muEq} shows, it is straightforward to shift the
resulting distance moduli to accommodate a new value for $\langle
M_{\rm corr}\rangle$. Unfortunately, we are not able to estimate
expansion velocities for five objects, and consequently not their
distances as well.

As mentioned in \S~\ref{INTR}, we can evaluate the accuracy of the SCM
from a comparison with EPM distances. For this purpose we employ EPM
distances recently calculated by \citet{Jon09}, which are summarized
in Table~\ref{TbEPM} along with our distances for the 11~objects in
common between SCM and EPM. \citet{Jon09} determine EPM distances with
two different sets of atmosphere models, both developed based on an
analytic approach to calculate dilution factors as a function of a
color temperature in order to correct the assumed blackbody
emission. Their EPM distances were derived from average dilution
factors and not from specific models for each supernova. The EPM
distances obtained using the models of \citet{Eas96} (E96; Column~2)
are $12 \pm 5$\% shorter than our SCM distances. On the other hand,
the EPM distances determined using the models of \citet{DeH05} (D05;
Column~3) are $40 \pm 10$\% greater than our SCM distances. Both
percentage shifts are calculated weighted by the uncertainty in the
EPM and SCM distances. These systematic differences can be seen in the
top panel of Figure~\ref{EPSC1}, and more clearly in the fractional
differences $(d_{\rm EPM}-d_{\rm SCM})/d_{\rm SCM}$ plotted in the
bottom panel.

The systematic differences among the two sets of EPM distances can be
solely attributed to the atmosphere models of E96 and D05. New
radiative transfer models of SNe~IIP are necessary to understand the
origin of this discrepancy. Besides the systematic errors in both EPM
implementations, we can gain an understanding of the internal
precision of EPM and SCM after removing the systematic differences and
bringing all distances to a common scale. For this purpose we correct
the EPM distances to the SCM distance scale by removing the percentage
shifts. The comparison is shown in Figure~\ref{EPSC2}; the distance
differences have dispersions of $\sim$~13\% and $\sim$~16\% using D05
and E96, respectively. This implies that SCM and EPM produce relative
distances with an internal precision $<13$--16\%. This agrees with the
dispersions of 0.25--0.30~mag seen in the HDs restricted to SNe with
$cz_{\rm CMB}>3000$~\kmpsec.

Further comparison between EPM and SCM can be made by taking the
distance determination to SN~1999em. While \citet{DeH06} give a value
of $11.5\pm1.0$~Mpc, our SCM returns $12.6\pm0.6$~Mpc. This agreement
has a significance better than 1$\sigma$, taking into account both
sources of error.

%% file: c5Disc.tex
\section{DISCUSSION} \label{DISC}

\subsection{Variations of the Extinction Law}

The reddening law that minimizes the dispersion in the $BVI$ HDs
($R_V=1.4\pm0.1$) turns out to be very different from the standard
Galactic extinction law \citep[$R_V=3.1$;][]{Car89}.  Recently,
\citet{Poz09} carried out a similar analysis using SNe~IIP, which
confirmed the low value obtained for $R_V$ presented by
\citet{Oli08}. There have also been several studies on this subject
using SNe~Ia.  Most recently, \citet{Fol10} solved for $R_V$ in the
same manner as our approach (i.e., by minimizing the dispersion in the
HD) and obtained a value of $R_V \approx 1.5$, remarkably consistent
with our value. Further evidence for low $R_{\lambda}$ values using
SNe~Ia were reported by others applying a similar procedure
\citep{Phi99,Alt04,Rei05,Hic09,Wan09}.

Other studies of dust reddening based on diverse methods were
performed for nearby galaxies giving values for $R_B$ ranging between
2.4 and 4.3 \citep{Rif90,DeP92,BrL91,Fin08}. The ratio of total to
selective absorption varies significantly over the range $R_V \approx
2.6-5.5$ even within our own Galaxy \citep{ClC88}.

A lower value of $R_V$ could be due to dust grains smaller than those
in our Galaxy, since for a given value of $A_V$ the $E(B-V)$ reddening
increases if the size of the grains decreases \citep{Dra03}. However,
it would be very unlikely that our Galaxy is so special.

Most probably we are dealing with normal dust grains in the
circumstellar medium, and therefore multiple scattering of photons
makes the reddening law steeper and $R_V$ smaller \citep{Wan05,Goo08}.
This effect should be opposite in the ultraviolet, hence it could be
further tested with photometry at such wavelengths.

\subsection{Internal Precision of the SCM}

The $BVI$ HDs have a scatter of $\sim$~0.3~mag, implying a precision
in individual distances of $\sim$~15\%. Part of this scatter might be
due to the peculiar velocities of the SN host galaxies, and the
intrinsic precision of SCM could be even better. In fact, when we
perform a comparison for SNe in common between SCM and EPM, the
distance differences are 13--16\% (after removing the systematic
difference among the SCM and EPM). This comparison is independent of
the SN host-galaxy redshifts and implies that the internal precision
in either of these two techniques must be lower than 13--16\%, since
these differences comprise the combined uncertainties of both
methods. This is an encouraging result; it implies that both EPM and
SCM can produce high-precision relative distances, thus offering a new
route to cosmological parameters.

\subsection{The LEV Relation}

We note that the dispersion in the HDs increases with wavelength. This
seems contrary to expectations given that (1) the extinction effects
decrease with wavelength, and (2) metallicity affects the $B$ band
more than the other optical bands. Perhaps the luminosity is a
function of not only velocity but also metallicity:
$M_V=f(\upsilon,\mbox{[Fe/H]})$.  If so, the velocity term in
Equations~(\ref{HDeq1}) and~(\ref{HDeq2}) might be removing
metallicity effects more efficiently at shorter wavelengths. This
could be tested with [Fe/H] measurements of SNe~IIP. On the other
hand, hydrodynamical simulations of SNe~IIP suggest that there is an
intrinsic scatter in the relation, which does not seem to be driven by
any parameter of the model. Given the evidence we have, we can only
claim a dependence of $L$ on $\upsilon$. The obvious physical
explanation is that the internal energy is correlated with the kinetic
energy, implying that the ratio of the internal energy to the kinetic
energy is approximately independent of the explosion energy. As
SN~1999br may suggest, underluminous events might (1) change the shape
of the LEV relation, or (2) require a different explosion mechanism.

%% file: c6Conc.tex
\section{CONCLUSIONS} \label{CONC}

We established a library of 196 SN~II optical spectra and developed a
code which allows us to correct the observed photometry for Galactic
extinction, $K$-terms, and host-galaxy extinction. We developed
fitting procedures to the light curves, color curves, and velocity
curves which allow us to precisely determine the transition time
between the plateau and the tail phases. The use of this parameter as
the time origin permitted us to line up the SNe to a common phase. The
additional benefit of these fits is the interpolation of magnitudes,
colors, and velocities over a wide range of epochs. The methodology
explained above yields the following conclusions.

\begin{enumerate}
\item The comparison between our color-based dereddening technique and
  the spectroscopic reddenings is satisfactory within the
  uncertainties of both techniques. This is particularly encouraging
  since our method uses late-time photometric information, while the
  other method uses completely independent early-time spectroscopic
  data.

\item Using our new sample of SNe we recover the luminosity versus
  expansion-velocity trend (LEV relation) previously found by
  \citet{HaP02}.

\item We construct $BVI$ Hubble diagrams using two sets of host-galaxy
  reddenings. We demonstrate that the extinction corrections based on
  $V-I$ color do a better job than the spectroscopic determinations,
  reaching dispersions of $\sim$~0.4~mag in the Hubble diagrams. This
  scatter is somewhat higher than the 0.35~mag found previously by
  \citet{Ha03b,Ham05}. A much smaller dispersion of 0.3~mag is
  achieved when we use $V-I$ colors to estimate reddening and allow
  $R_V$ to vary. We obtain $R_V=1.4\pm0.1$, much smaller than the
  Galactic value of 3.1. The low value of $R_V$ can be explained by a
  different nature of the dust grains in the host galaxies of SNe~IIP
  or by scattering of light by circumstellar dust clouds
  \citep{Wan05,Goo08}.

\item A clear correlation is found among residuals of $BVI$ HDs. After
  removing the correlation, the rms in the HDs decreases to 6--9\% in
  distance, comparable to the precision achieved with SNe~Ia.

\item After calibrating our Hubble diagrams with Cepheid distances to
  SN~1999em and SN~2004dj, our $BVI$ photometry leads to a Hubble
  constant in the range 62--105~\dimho; this agrees with most of the
  values obtained by means of diverse modern techniques.

\item Finally, we calculate the distance moduli to our SNe, and make a
  comparison against EPM distances from \citet{Jon09}. The 11~SNe in
  common show a systematic difference in distance between EPM and SCM,
  depending on the atmosphere models employed by EPM. Correcting for
  these shifts we bring the EPM distances to the SCM distance scale,
  from which we measure a dispersion of 13--16\%. This spread reflects
  the combined internal precision of EPM and SCM. Therefore, the
  internal precision in either of these two techniques must be
  $<$~13--16\%.
\end{enumerate}

This analysis confirms the utility of SNe~IIP as cosmological probes,
providing strong encouragement to future high-redshift studies. We
find that one can determine relative distances from SNe~IIP with a
precision of 15\% or better. This uncertainty could be further reduced
by including more SNe in the Hubble flow.  In its current form the SCM
requires both photometric and spectroscopic data. Since the latter are
expensive to obtain (especially at high redshifts), it would be
desirable to look for a photometric observable as a luminosity
indicator instead of the expansion velocities.

%% file: zAckn.tex
\acknowledgments 

We thank Luc Dessart, who generously provided the reddening estimates
calculated by means of spectral fitting and the models plotted in
Figure~\ref{FgSpec}. F.O.E. also thanks him for stimulating
discussions, as well as for accurate comments and suggestions. We are
grateful to D.~Leonard, J.~Vink\'o, D.~Sahu, and A.~Pastorello for
providing digital spectra of SN~1999gi, SN~2004dj, SN~2004et, and
SN~2005cs, respectively. F.O.E., M.H., G.P., and J.M. acknowledge
support from the Millennium Center for Supernova Science through grant
Pas06--045--F funded by ``Programa Bicentenario de Ciencia y
Tecnolog\'ia de CONICYT'' and ``Programa Iniciativa Cient\'ifica
Milenio de MIDEPLAN,'' as well as support provided by Centro de
Astrof\'{\i}sica FONDAP 15010003 and by Fondecyt through grant 1060808
from the Center of Excellence in Astrophysics and Associated
Technologies (PFB 06). The supernova research of A.V.F.'s group at
U.C. Berkeley has been financed by NSF grants AST--0607485 and
AST-0908886, as well as by the TABASGO Foundation. Supernova research
at the Harvard College Observatory has been supported by NSF grant
AST--0606772. G.P. acknowledges partial support from Comit\'e Mixto
ESO-GOBIERNO DE CHILE. This paper is based in part on observations
taken at the Cerro Tololo Inter-American Observatory, National Optical
Astronomy Observatory, which is operated by the Association of
Universities for Research in Astronomy, Inc. (AURA) under cooperative
agreement with the NSF. Partly based on observations collected at the
European Southern Observatory, Chile, in the course of program
163.H-0285.

%% file: zFaci.tex
{\it Facilities:} \facility{CTIO:0.9m (SITe No. 3 imaging CCD)},
\facility{CTIO:1.5m (CCD, CSPEC)}, \facility{YALO (ANDICAM, 2DF)},
\facility{Blanco (CSPEC, 2DF, CCD)}, \facility{Swope (SITe No. 3
imaging CCD)}, \facility{du Pont (Tek No. 5 imaging CCD, WFCCD,
MODSPEC, 2DF)}, \facility{Magellan:Baade (LDSS2 imaging spectrograph,
Boller \& Chivens spectrograph)}, \facility{Magellan:Clay (LDSS2
imaging spectrograph)}, \facility{ESO:1.52m (IDS)}, \facility{Danish
1.54m Telescope (DFOSC)}, \facility{Max Plank:2.2m (EFOSC2)},
\facility{NTT (EMMI)}, \facility{ESO:3.6m (EFOSC)},
\facility{SO:Kuiper (CCD)}, \facility{Bok (B\&C)}, \facility{Lick:3.0m
(Kast)}, \facility{FLWO:1.5m (FAST)}

%% file: 0_online/f31LCfit.tex
\begin{figure}[p]
\begin{center}
\includegraphics[angle=0,scale=0.8]{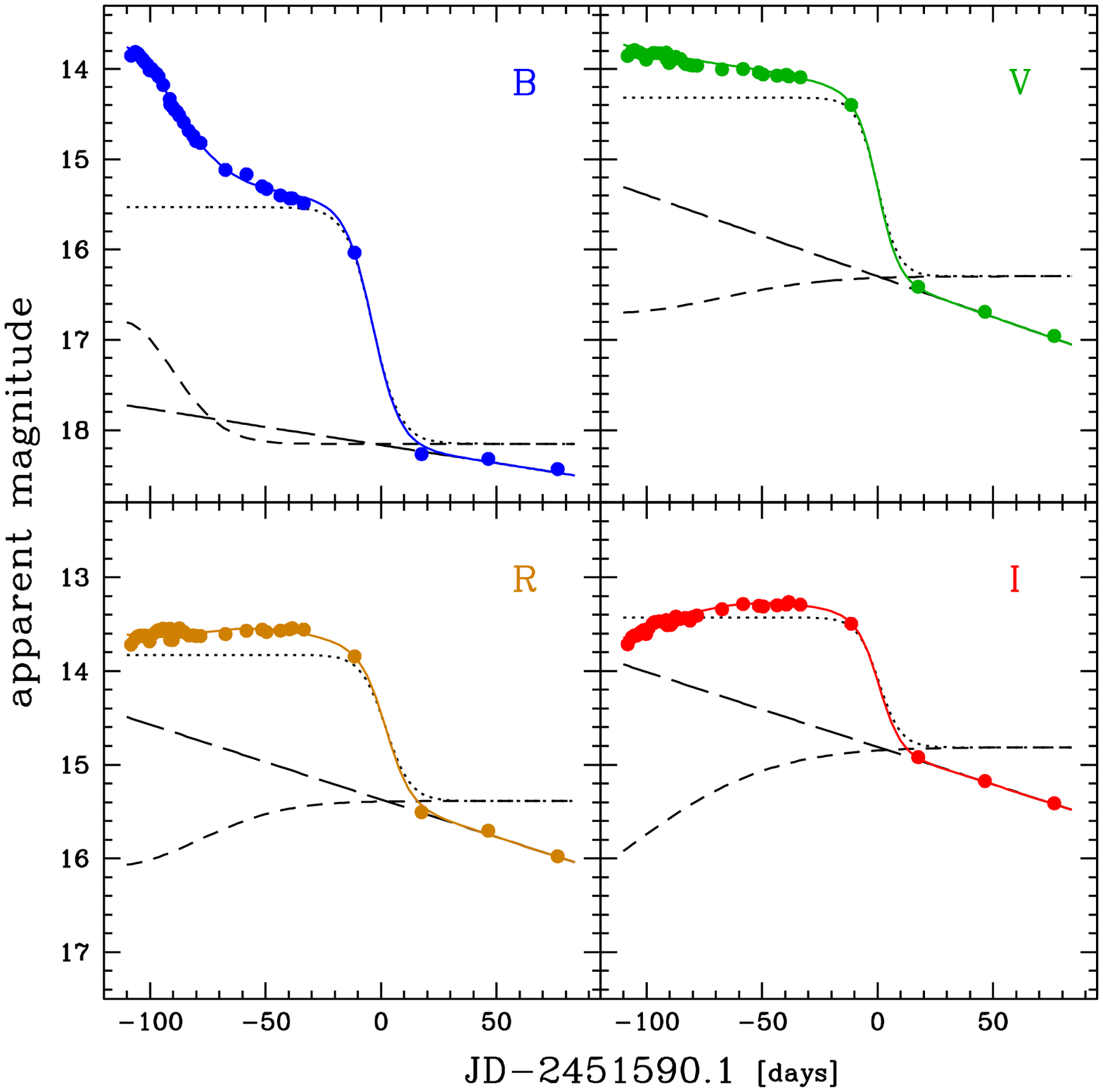}
\caption{$BVRI$ light curves of SN~1999em. $BVRI$ magnitudes are
  respectively shown in blue (top-left), green (top-right), brown
  (bottom-left), and red (bottom-right). The colored solid line
  corresponds to the analytic fit. The black lines represent the
  decomposition of the main function into its three components: the
  Fermi-Dirac function (dotted line), a straight line (long-dashed
  line), and a Gaussian function (short-dashed line). The origin of
  the time axis corresponds to $t_{PT}$ in Julian days for
  SN~1999em. Since the instrumental error bars are smaller than the
  symbols, they are not shown. \label{FgLCfit}}
\end{center}
\end{figure}

%% file: 0_online/f32Colr.tex
\begin{figure}[p]
\begin{center}
\includegraphics[angle=0,scale=0.8]{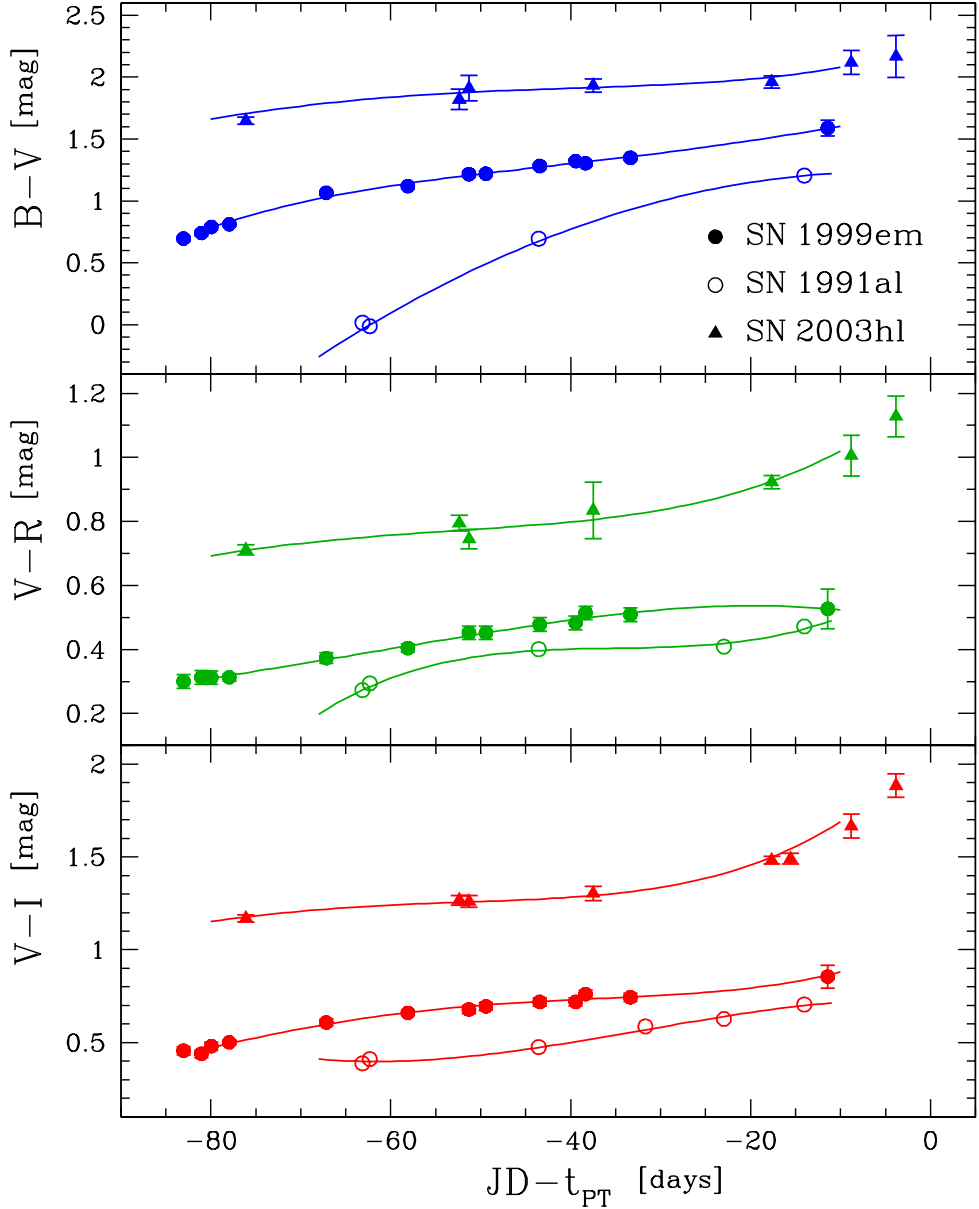}
\caption{$B-V$, $V-R$, and $V-I$ color curves of SN~1999em (filled
  circles), SN~1991al (open circles), and SN~2003hl (filled triangles)
  corrected for the $A_{\rm G}$ and $K$-terms. This comparison
  demonstrates that each SN displays a different color evolution,
  precluding the determination of color excesses from a simple color
  offset after correcting for the $A_{\rm G}$ and $K$-terms. For
  SN~1991al the error bars are always smaller than the symbols and
  therefore not plotted. \label{FgColr}}
\end{center}
\end{figure}

%% file: 0_online/f33Vels.tex
\begin{figure}[p]
\begin{center}
\includegraphics[angle=0,scale=0.8]{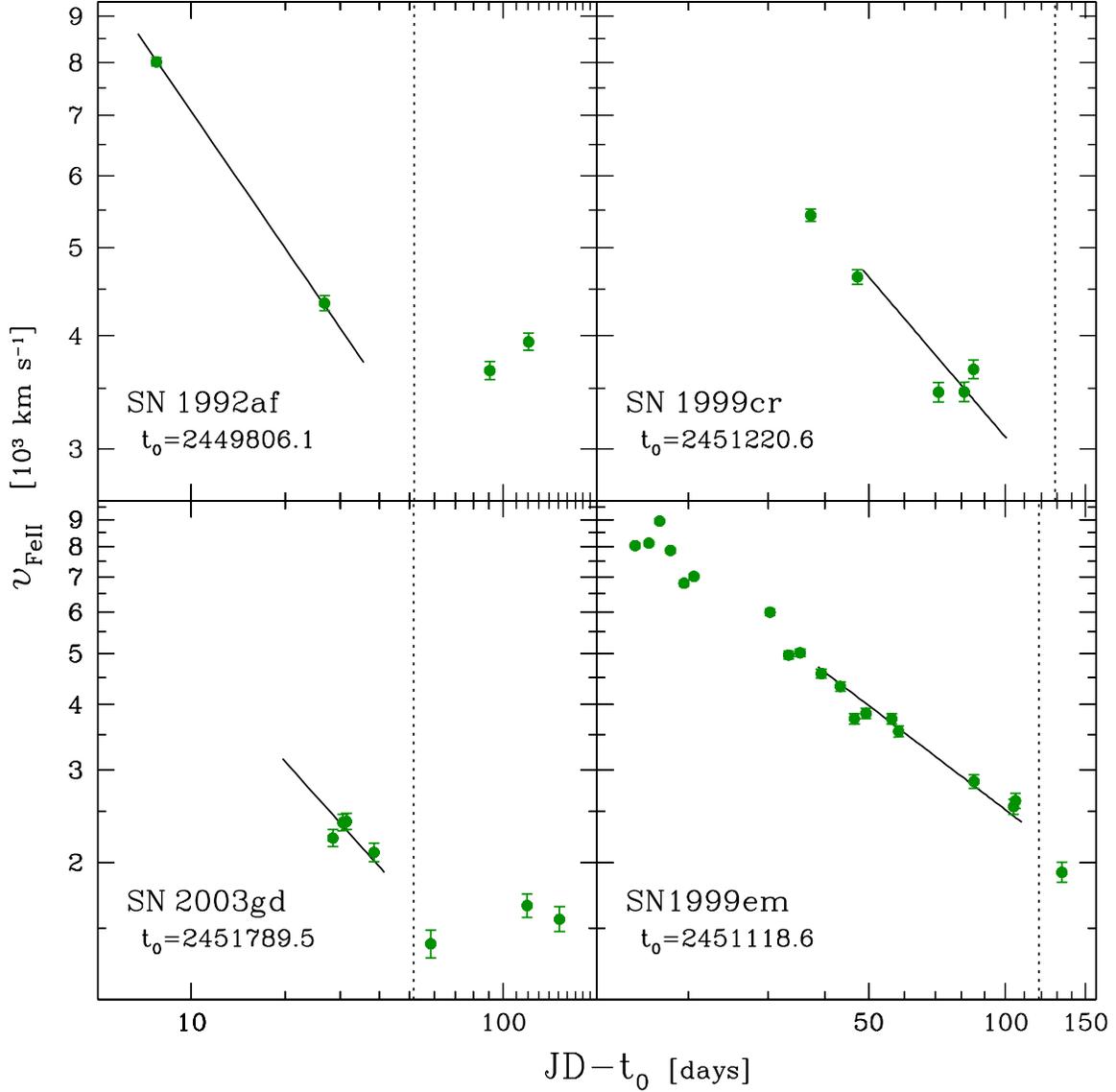}
\caption{Four expansion velocity curves representative of the SN
  sample measured by means of the \ion{Fe}{2}~$\lambda$5169 line
  profile. While solid lines correspond to a power-law fit
  (equation~\ref{vel_eq}), dashed lines show the transition time. The
  $t_0$ parameter of the fit was choosen as the time origin due to
  plotting restrictions. For SN~1992af (top-left panel) we fit only
  two velocities (see \S~\ref{exp} for more details).  The top-right
  panel shows a 15-day extrapolation beyond the last data point for
  SN~1999cr. The bottom-left panel shows a 10-day extrapolation for
  SN~2003gd prior to the first data point. SN~1999em velocities are
  shown in the bottom-right panel as an example of good
  sampling.\label{FgVels}}
\end{center}
\end{figure}

%% file: 0_online/f34Spec.tex
\begin{figure}[p]
\begin{center}
\includegraphics[angle=0,scale=0.8]{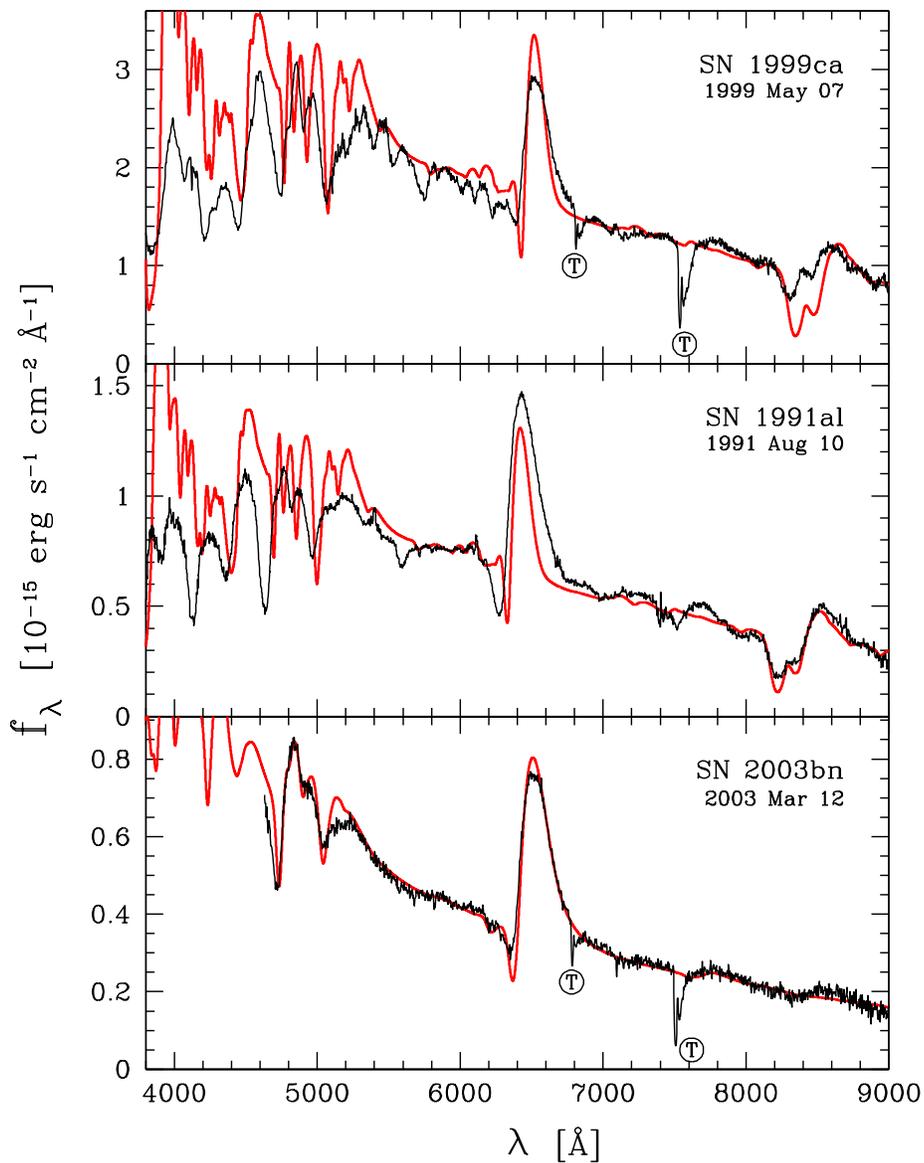}
\caption{Type~IIP SN atmosphere models (wide red line) fitted to our
  spectra (black line) in the rest frame. The top panel shows an
  example of an unsatisfactory fit to one of our late-time
  spectra. The middle panel shows a better fit to an early-time
  spectrum of SN~1991al. A much better fit is achieved for an
  early-time spectrum of SN~2003bl as shown in the bottom
  panel. Strong telluric lines are marked. \label{FgSpec}}
\end{center}
\end{figure}

%% file: 0_online/f35VIVR.tex
\begin{figure}[p]
\begin{center}
\includegraphics[angle=0,scale=0.7]{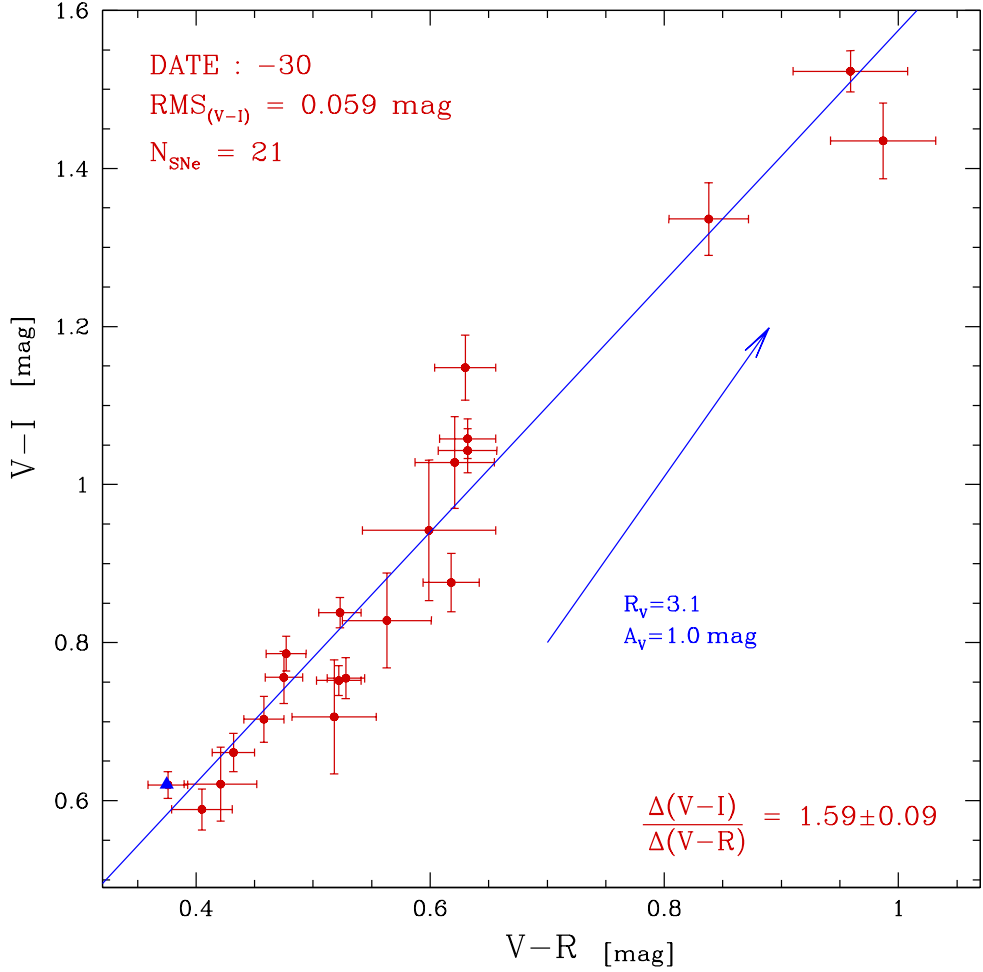}
\caption{$V-I$ versus $V-R$ diagram for 21~SNe~IIP having $VRI$
  photometry corrected for the $A_{\rm G}$ and $K$-terms. The blue
  line is a least-squares fit to the data, with a slope of
  $1.59\pm0.09$. The blue arrow has a slope of~2.12 and corresponds to
  the reddening vector for $A_V=1.0$~mag and a standard Galactic
  extinction law ($R_V=3.1$). The blue triangle shows the one SN of
  this subsample consistent with zero extinction having $R$-band
  photometry. \label{FgVIVR}}
\end{center}
\end{figure}

%% file: 0_online/f36VIBV.tex
\begin{figure}[p]
\begin{center}
\includegraphics[angle=0,scale=0.7]{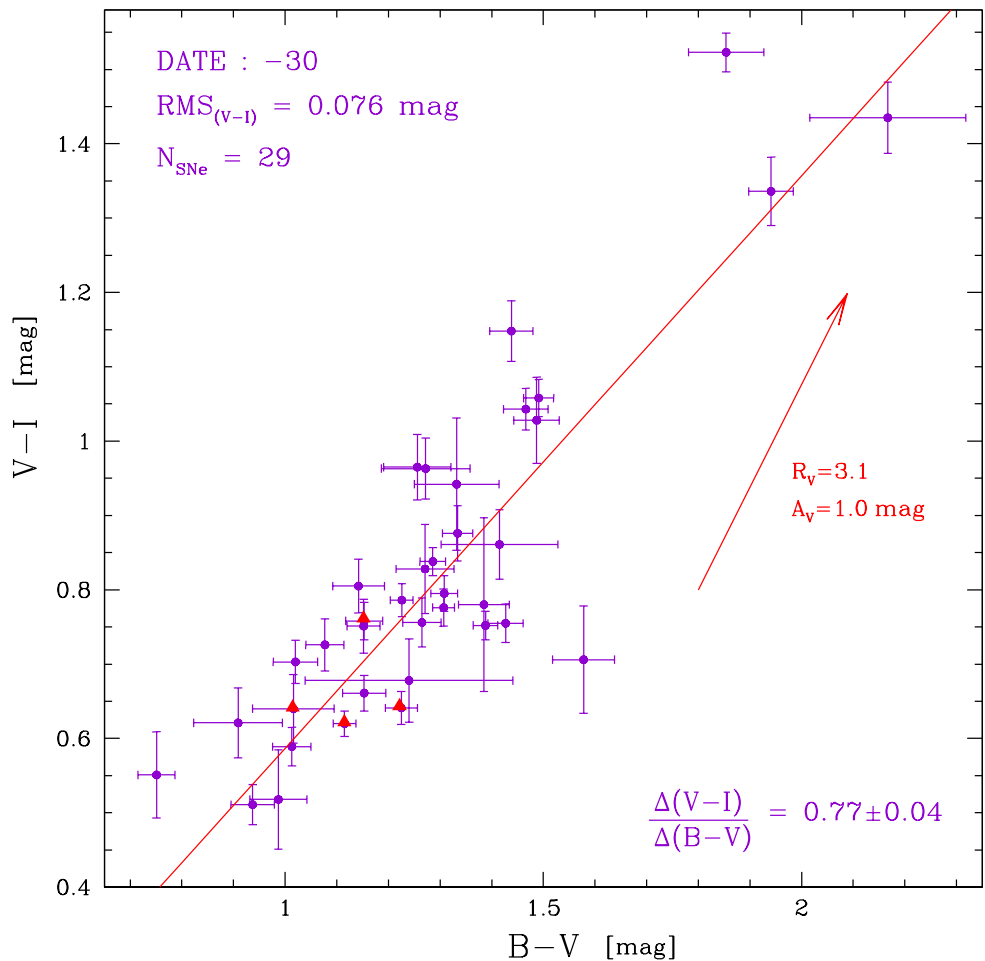}
\caption{$V-I$ versus $B-V$ diagram for 29~SNe~IIP having $BVI$
  photometry corrected for the $A_{\rm G}$ and $K$-terms. The red line
  is a least-squares fit to the data, with a slope of
  $0.77\pm0.04$. The red arrow has a slope of~1.38 and corresponds to
  the reddening vector for $A_V=1.0$~mag and a standard Galactic
  extinction law ($R_V=3.1$). The red triangles show the four SNe
  consistent with zero host-galaxy extinction. \label{FgVIBV}}
\end{center}
\end{figure}

%% file: 0_online/f401AviAspc.tex
\begin{figure}[p]
\begin{center}
\includegraphics[angle=0,scale=0.7]{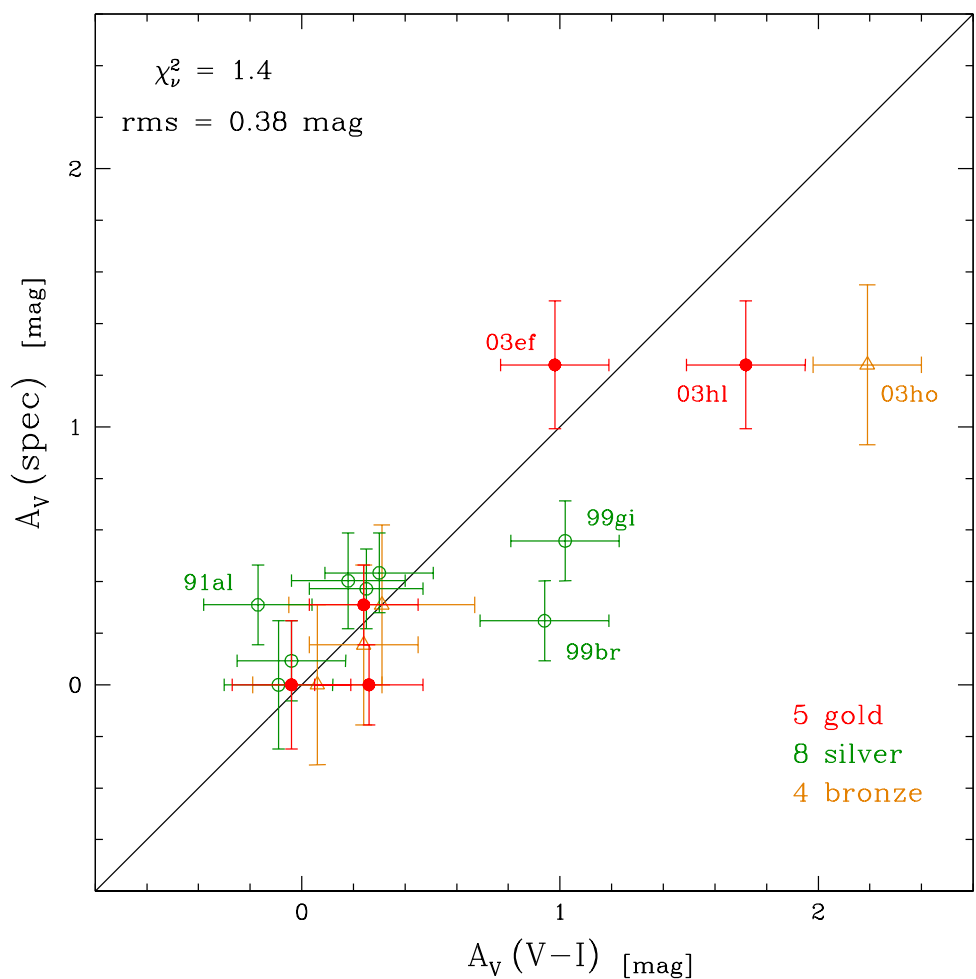}
\caption{Comparison between two dereddening techniques via
  $A_V$~(mag): the spectrum-fitting method (ordinate) and our
  technique (abscissa; \S~\ref{aho}). The figure shows three
  subclasses defined by the quality of the spectroscopic data used:
  the {\it gold}, {\it silver}, and {\it bronze} samples (see
  \S~\ref{dtc}).  The solid black line is fixed to show a slope of
  unity. \label{FgRd1}}
\end{center}
\end{figure}

%% file: 0_online/f402AviAnad.tex
\begin{figure}[p]
\begin{center}
\includegraphics[angle=0,scale=0.7]{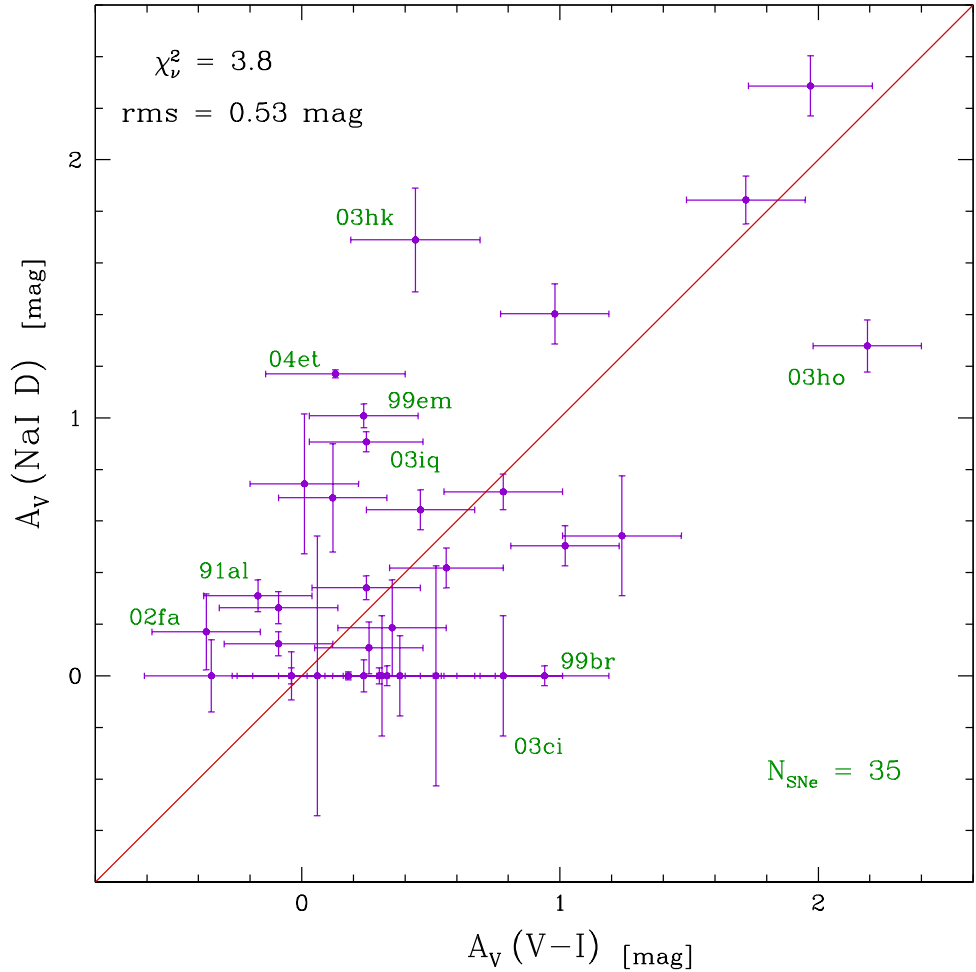}
\caption{Comparison between two dereddening techniques via
  $A_V$~(mag): the \nad\ interstellar line (ordinate) and our
  technique (abscissa; \S~\ref{aho}). In each case, the equivalent
  width of the \nad\ interstellar line is transformed into a visual
  extinction according to the calibration given by
  \citet{Bar90}. \label{FgNad}}
\end{center}
\end{figure}

%% file: 0_online/f403LmVe.tex
\begin{figure}[p]
\begin{center}
\includegraphics[angle=0,scale=0.75]{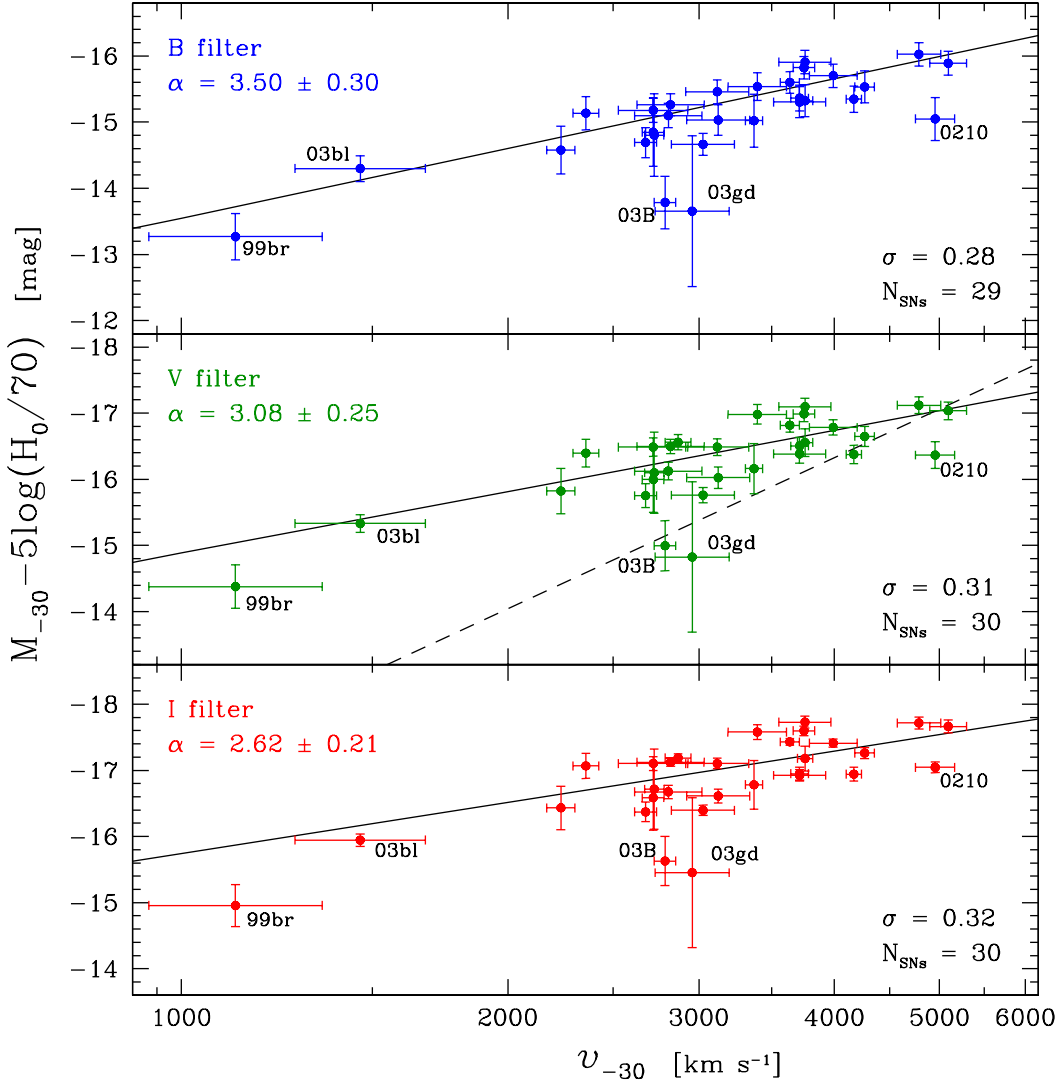}
\caption{$BVI$-band absolute magnitudes (ordinate) against the
  \ion{Fe}{2} expansion velocity (abscissa) for 29--30~SNe.  We use
  magnitudes corrected for Galactic extinction, $K$-terms, and
  host-galaxy extinction together with a value of the Hubble constant
  of 70~\dimho. The dashed line in the middle panel represents the LEV
  relation for the $V$ band found by \citet{HaP02}, obtained from
  magnitudes and velocities measured at day~50 past the explosion
  (approximately day~$-60$ on our own time scale).\label{FgLmVe}}
\end{center}
\end{figure}

%% file: 0_online/f404HDbA.tex
\begin{figure}[p]
\begin{center}
\includegraphics[angle=0,scale=0.8]{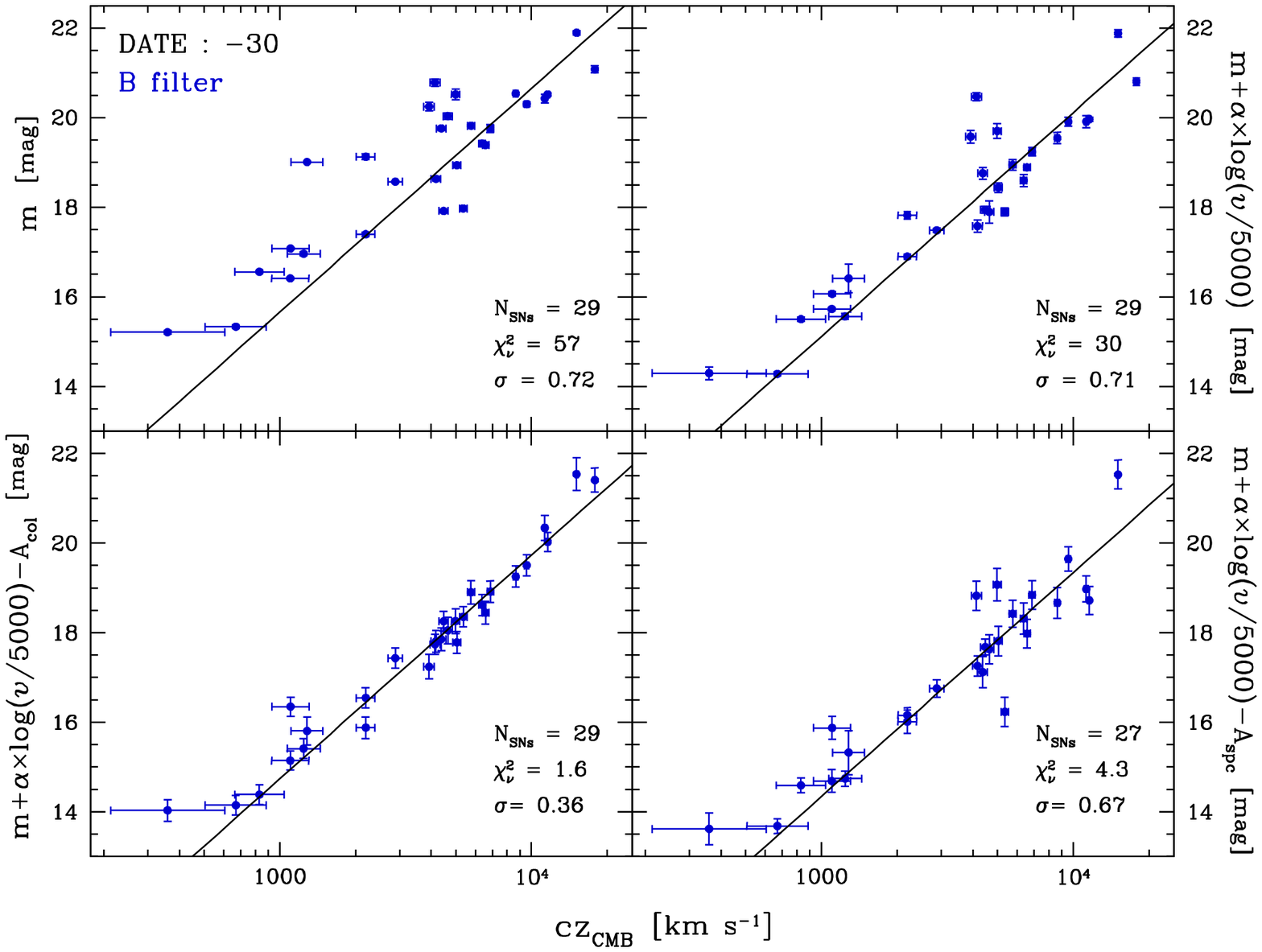}
\caption{$B$-band Hubble diagrams. The top-left panel shows magnitudes
  interpolated to day~$-30$ previously corrected for Galactic
  extinction and $K$-terms; the top-right panel shows magnitudes
  additionally corrected for expansion velocities; the bottom-left
  panel includes further corrections for color-based host-galaxy
  extinction (using the $V-I$ color calibration given in
  \S~\ref{aho}); and the bottom-right panel replaces the color-based
  extinction for the spectroscopic reddenings. The horizontal bars in
  each panel correspond to an adopted uncertainty of 187~\kmpsec\ in
  the CMB redshift.
  \label{HDB}}
\end{center}
\end{figure}

%% file: 0_online/f405HDvA.tex
\begin{figure}[p]
\begin{center}
\includegraphics[angle=0,scale=0.8]{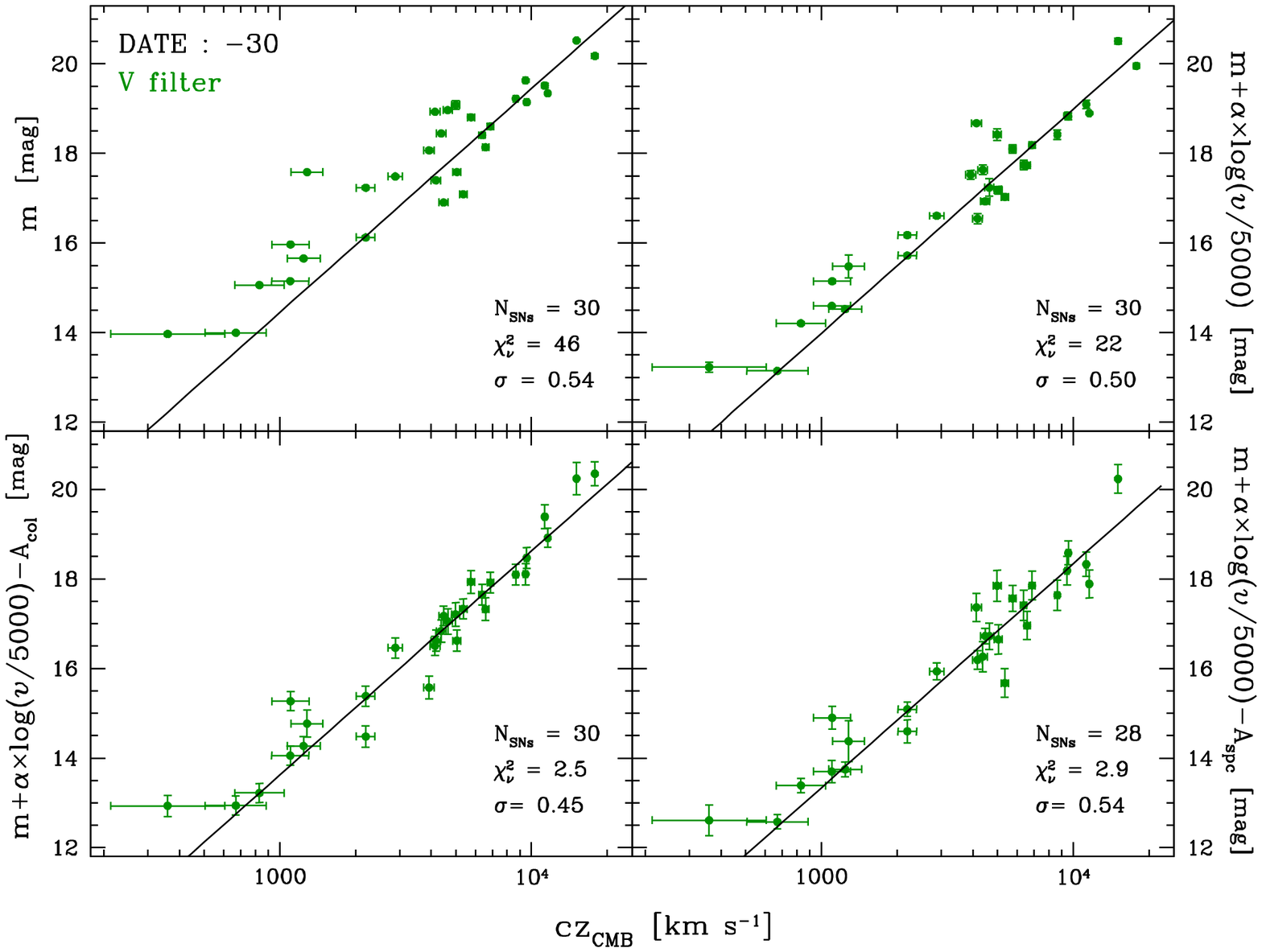}
\caption{$V$-band Hubble diagrams. The top-left panel shows magnitudes
  interpolated to day~$-30$ previously corrected for Galactic
  extinction and $K$-terms; the top-right panel shows magnitudes
  additionally corrected for expansion velocities; the bottom-left
  panel includes further corrections for color-based host-galaxy
  extinction (using the $V-I$ color calibration given in
  \S~\ref{aho}); and the bottom-right panel replaces the color-based
  extinction for the spectroscopic reddenings. The horizontal bars in
  each panel correspond to an adopted uncertainty of 187~\kmpsec\ in
  the CMB redshift.
  \label{HDV}}
\end{center}
\end{figure}

%% file: 0_online/f406HDiA.tex
\begin{figure}[p]
\begin{center}
\includegraphics[angle=0,scale=0.8]{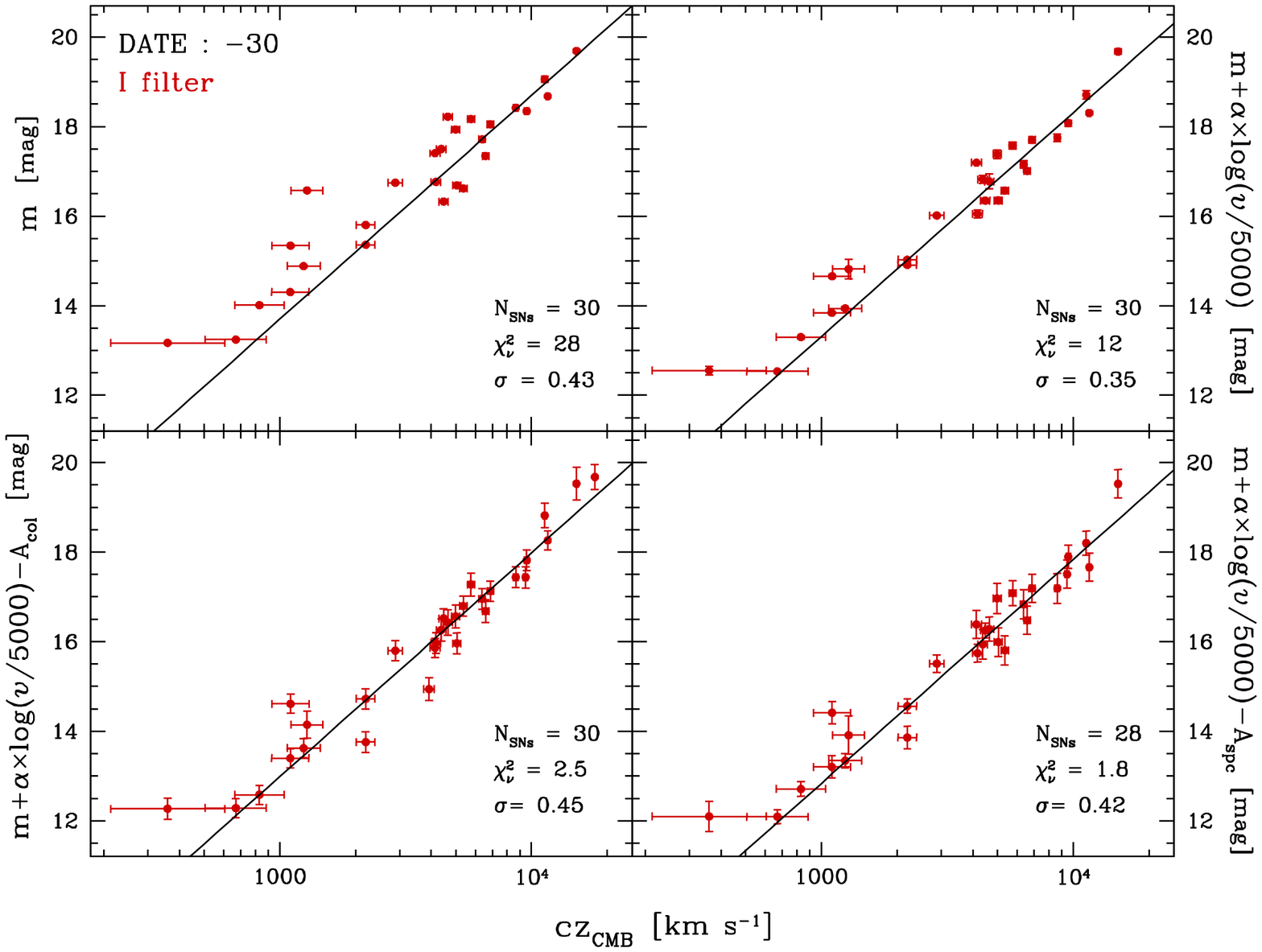}
\caption{$I$-band Hubble diagrams. The top-left panel shows magnitudes
  interpolated to day~$-30$ previously corrected for Galactic
  extinction and $K$-terms; the top-right panel shows magnitudes
  additionally corrected for expansion velocities; the bottom-left
  panel includes further corrections for color-based host-galaxy
  extinction (using the $V-I$ color calibration given in
  \S~\ref{aho}); and the bottom-right panel replaces the color-based
  extinction for the spectroscopic reddenings. The horizontal bars in
  each panel correspond to an adopted uncertainty of 187~\kmpsec\ in
  the CMB redshift.
  \label{HDI}}
\end{center}
\end{figure}

%% file: 0_online/f407HDbviRv.tex
\begin{figure}[p]
\begin{center}
\includegraphics[angle=0,scale=0.8]{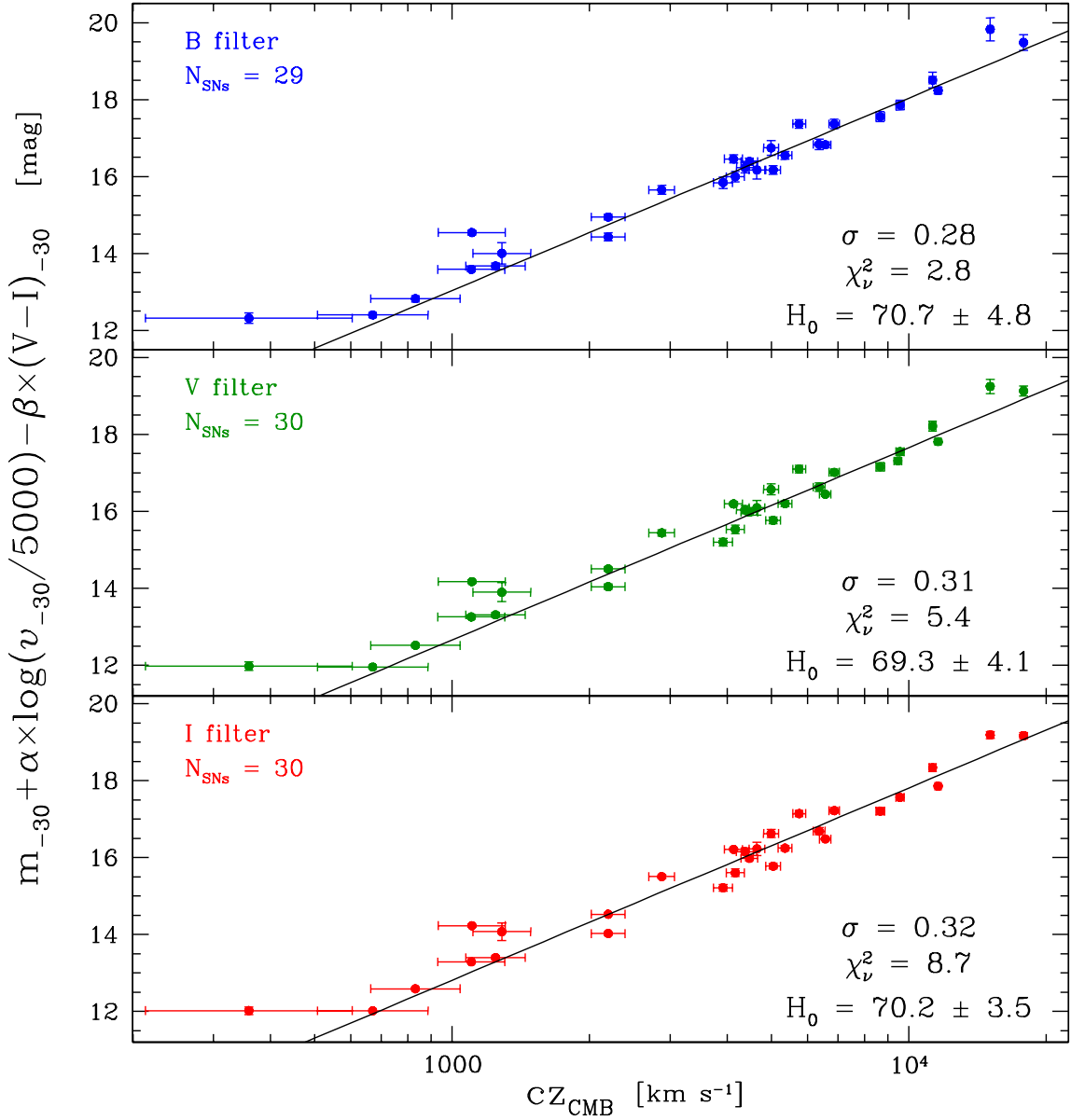}
\caption{$BVI$ Hubble diagrams leaving $R_V$ as a free parameter. In
  the top, middle, and bottom panels we show the HDs using $BVI$-band
  magnitudes of 29, 30, and 30~SNe, respectively. The horizontal bars in
  each panel correspond to an adopted uncertainty of 187~\kmpsec\ in
  the CMB redshift.
  \label{HDBVI}}
\end{center}
\end{figure}

%% file: 0_online/f408RvBeta.tex
\begin{figure}[p]
\begin{center}
\includegraphics[angle=0,scale=0.7]{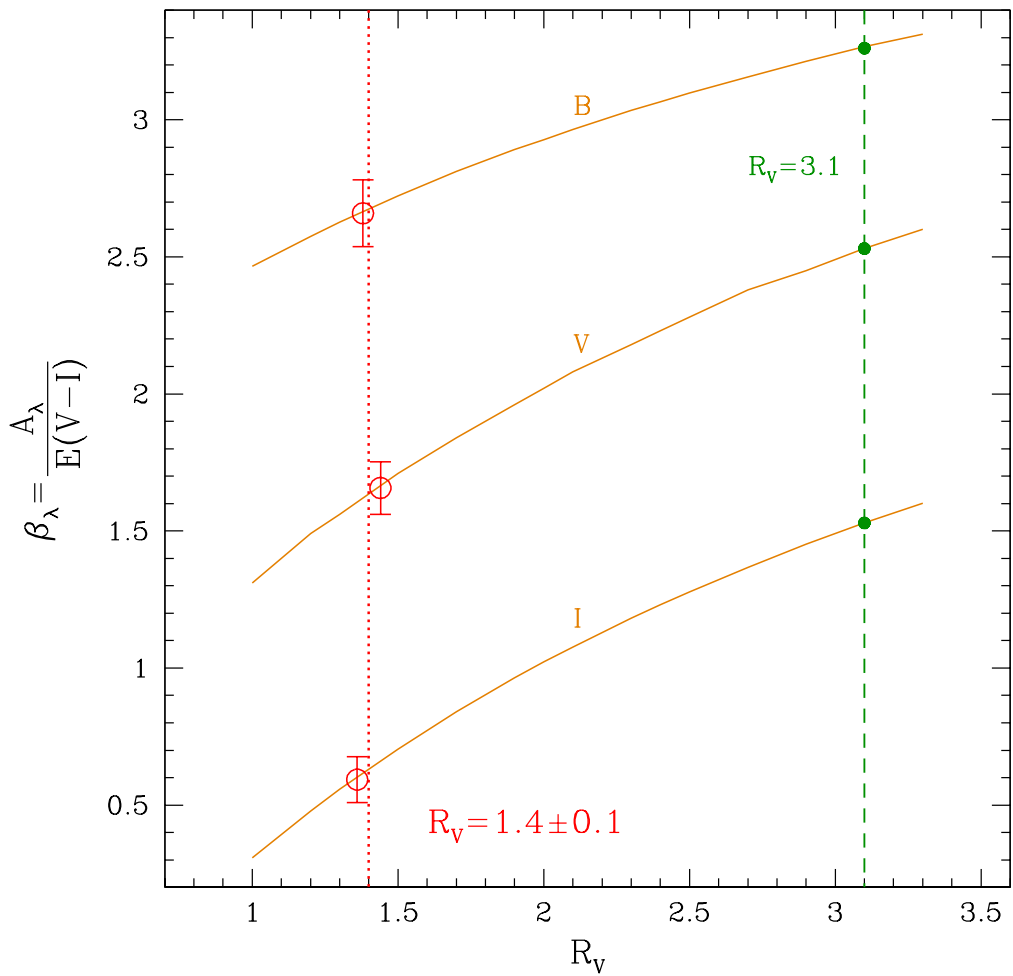}
\caption{$R_V$ versus $\beta$. The solid lines are computed from our
  library of SN~II spectra for each of the $BVI$ bands. The $\beta$
  parameters derived from the Hubble diagrams in Figure~\ref{HDBVI}
  are shown with red open circles for each of the $BVI$ bands. The red
  dotted line shows the $R_V$-weighted mean for the $BVI$ values of
  $\beta$.  The values for the standard Galactic extinction law are
  also shown with green dots. \label{FgRv}}
\end{center}
\end{figure}

%% file: 0_online/f409resM.tex
\begin{figure}[p]
\begin{center}
\includegraphics[angle=0,scale=0.8]{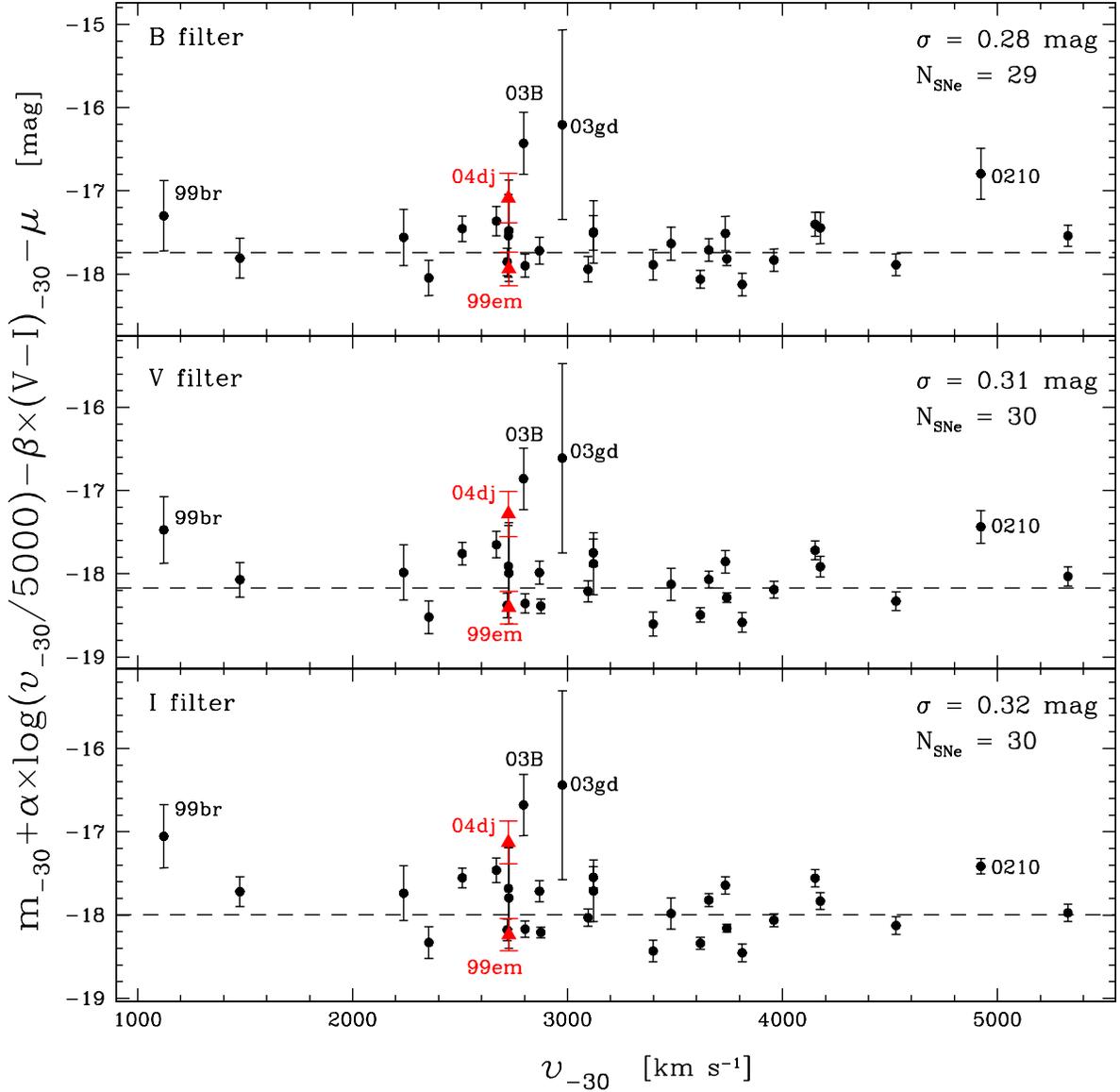}
\caption{$BVI$-corrected absolute magnitudes (ordinate) against the
  \ion{Fe}{2} expansion velocity (abscissa) for 29--30~SNe (black
  dots). We use the $H_0$ values of Table~\ref{TbHo} and the CMB
  redshifts to compute the distance moduli. The dashed lines show the
  mean corrected absolute magnitude for the black data points. The red
  triangles are the two calibrating SNe, whose corrected absolute
  magnitudes were calculated using their Cepheid distances (see
  Table~\ref{TbHo}). Note that, as expected, the two calibrating SNe
  fall on each side of the corrected absolute magnitude
  distributions. \label{Mcor}}
\end{center}
\end{figure}

%% file: 0_online/f410resCor.tex
\begin{figure}[p]
\begin{center}
\includegraphics[angle=0,scale=0.8]{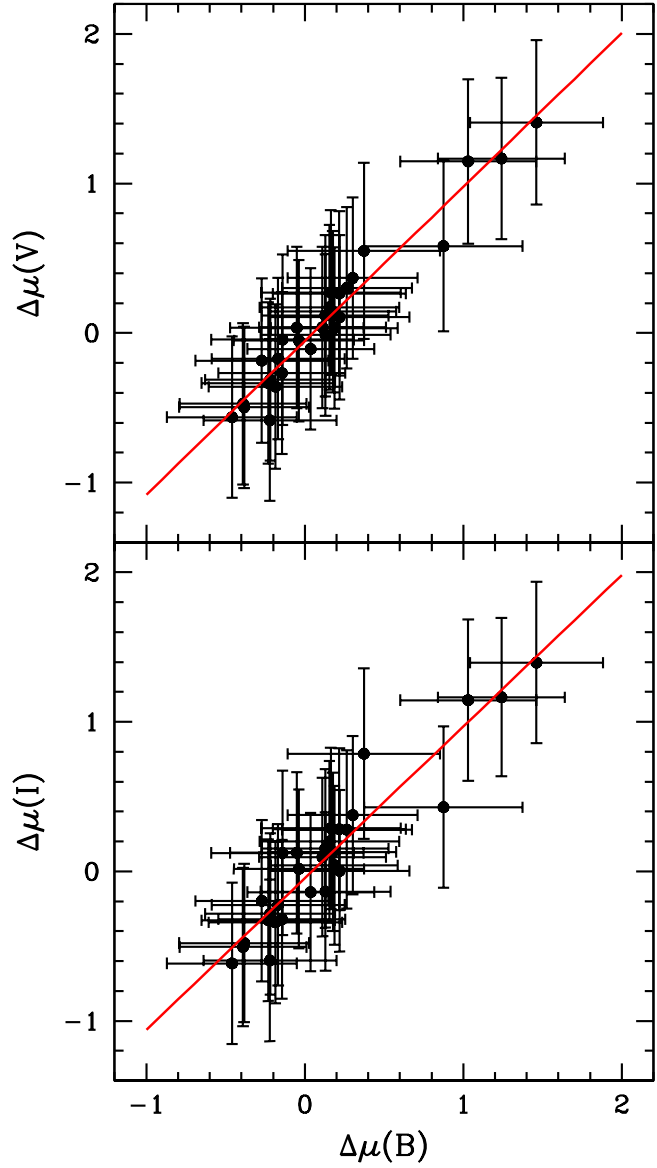}
\caption{Residuals from the $BVI$ HDs. The correlation between the
  $V$-filter and $I$-filter residuals against the $B$-filter residuals
  is shown in the upper and lower panels, respectively. The red lines
  represent the fit from which the scatter is measured.\label{resCor}}
\end{center}
\end{figure}

%% file: 0_online/f410EPMSCM.tex
\begin{figure}[p]
\begin{center}
\includegraphics[angle=0,scale=0.8]{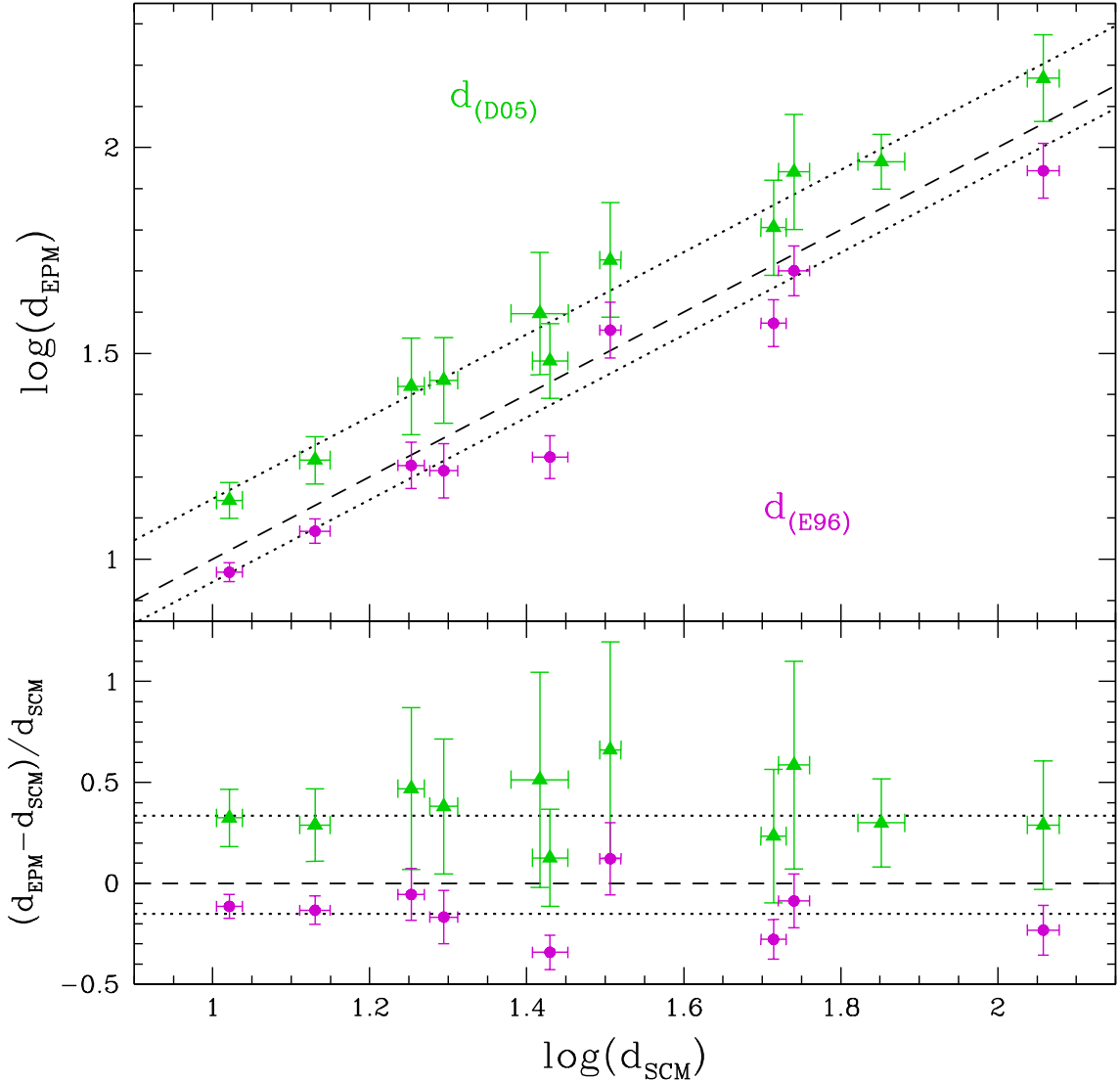}
\caption{Direct comparison between SCM and EPM distances in Mpc
  calculated by \citet{Jon09}. In magenta are shown the EPM distances
  computed from the atmosphere models of E96, while the green
  triangles are EPM distances obtained from the D05 models. The bottom
  panel shows the fractional differences between both techniques
  against $\log(d_{\rm SCM})$. In both panels the dotted lines trace
  the systematic shifts of the EPM distances with respect to the SCM
  distances, $\sim$~40\% and $\sim$~12\% using D05 and E96 atmosphere
  models, respectively. The dashed line in the top panel is fixed to
  show a slope of unity. \label{EPSC1}}
\end{center}
\end{figure}

%% file: 0_online/f411EPMSCM.tex
\begin{figure}[p]
\begin{center}
\includegraphics[angle=0,scale=0.8]{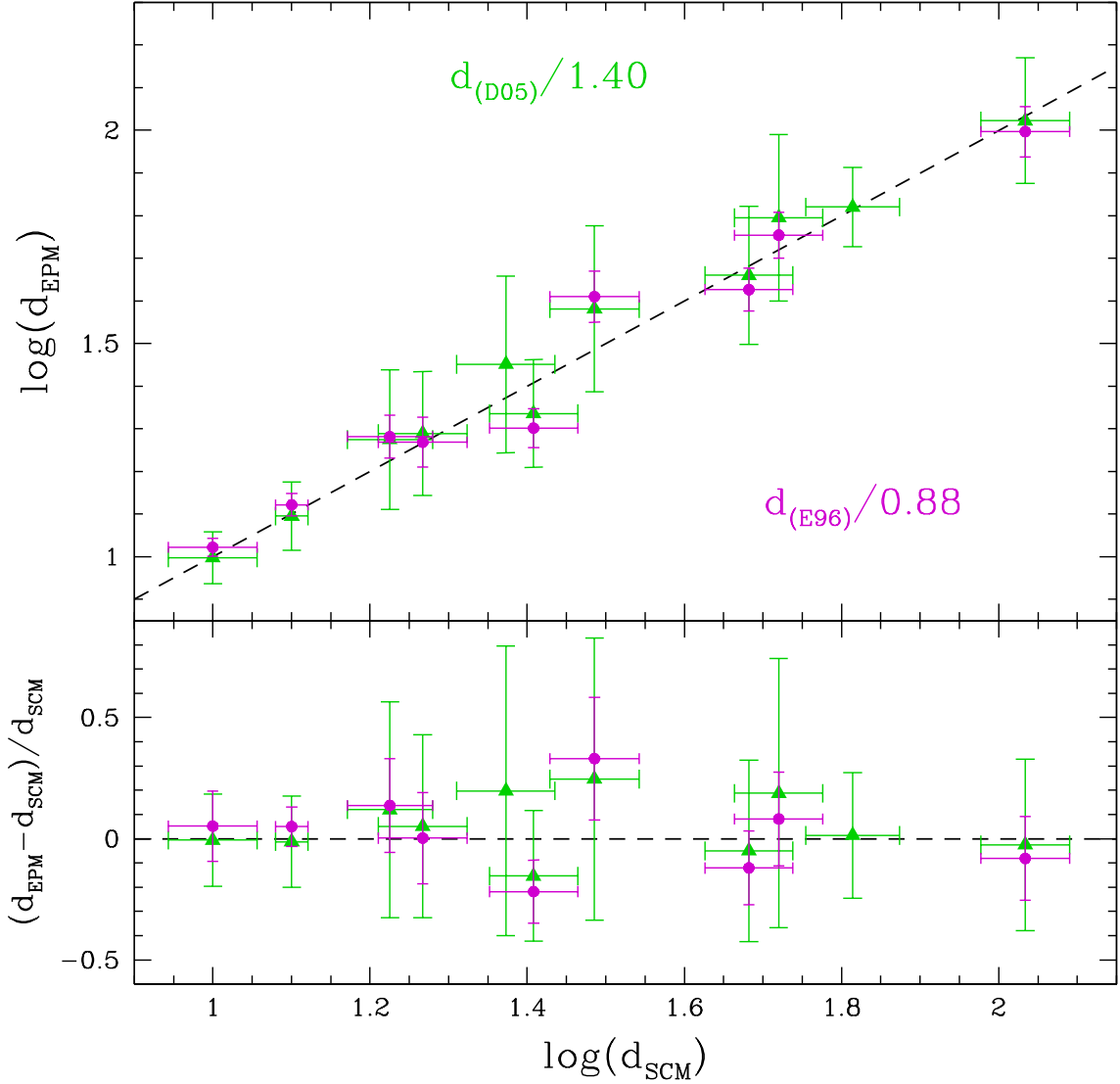}
\caption{Similar to Figure~\ref{EPSC1}, after removing systematic
  differences by bringing the EPM distances to the SCM scale (all in
  Mpc). Only random differences between EPM and SCM distances
  remain. For the E96 case the discrepancies have a spread of 16\%,
  while for the D05 case the scatter is 13\%. The dashed line in the
  top panel is fixed to show a slope of unity.
\label{EPSC2}}
\end{center}
\end{figure}

%% file: t21Teles.tex
\begin{table}[p]
\begin{center}
{\scshape \caption{Telescopes and Instruments}\label{TbIns}}
\vspace{3mm}
\begin{tabular}{lcc}
\hline\hline
Telescope      &Instrument &Phot/Spec \\
\hline
CTIO 0.9 m     &  CCD      &  P   \\
YALO 1.0 m     &  ANDICAM  &  P   \\
YALO 1.0 m     &  2DF      &  S   \\
CTIO 1.5 m     &  CCD      &  P   \\
CTIO 1.5 m     &  CSPEC    &  S   \\
Blanco 4.0 m   &  CSPEC    &  S   \\
Blanco 4.0 m   &  2DF      &  S   \\
Blanco 4.0 m   &  CCD      &  P   \\
Swope 1.0 m    &  CCD      &  P   \\
du Pont 2.5 m  &  WFCCD    &  P/S \\ 
du Pont 2.5 m  &  MODSPEC  &  S   \\
du Pont 2.5 m  &  2DF      &  S   \\
du Pont 2.5 m  &  CCD      &  P   \\
Baade 6.5 m    &  LDSS2    &  P/S \\
Baade 6.5 m    &  B\&C     &  S   \\
Clay 6.5 m     &  LDSS2    &  P/S \\
ESO 1.5 m      &  IDS      &  S   \\
Danish 1.5 m   &  DFOSC    &  P/S \\
MPI/ESO 2.2 m  &  EFOSC2   &  S   \\
NTT 3.6  m     &  EMMI     &  S   \\
ESO 3.6 m      &  EFOSC    &  S   \\
Kuiper 1.5 m   &  CCD      &  P   \\
Bok 2.2 m      &  B\&C     &  S   \\
Lick 3.0 m     &  Kast     &  S   \\
Whipple 1.5 m  &  FAST     &  S   \\
\hline
\end{tabular}

\tablecomments{Whether the instrument was used for photometry (P),
  spectroscopy (S), or both (P/S) is listed in Column~3.}

\normalsize
\end{center}
\end{table}

%% file: t22List.tex
\begin{table}[p]
\begin{center}
{\scshape \caption{Supernova Sample}\label{Tb1}}
\scriptsize
\vspace{2mm}
\begin{tabular}{lcrrcccc}
\hline\hline
SN Name &Host Galaxy      &\multicolumn{1}{c}{$\alpha$(J2000)}   &\multicolumn{1}{c}{$\delta$(J2000)}        &$z_{\rm host}$\tablenotemark{a} &(s)\tablenotemark{b} &$E(B-V)_{\rm Gal}$\tablenotemark{c} &References  \\
        &                 &\multicolumn{1}{l}{[ h \ m \ s ]}     &\multicolumn{1}{c}{[ $^\circ$ \ ' \ '' ]}  &                                &                     &[mag]                               &          \\
\hline                                                                                          
1991al  &LEDA 140858      &19 42 24.00  &$-55$ 06 23.0  &0.01525  &HP02	&0.051    &1       \\   
1992af  &ESO 340-G038     &20 30 40.20  &$-42$ 18 35.0  &0.01847  &NED	&0.052    &1       \\   
1992am  &MCG--01--04--039 &01 25 02.70  &$-04$ 39 01.0  &0.04773  &NED	&0.049    &1       \\   
1992ba  &NGC 2082         &05 41 47.10  &$-64$ 18 01.0  &0.00395  &NED	&0.058    &1       \\   
1993A   &anonymous        &07 39 17.30  &$-62$ 03 14.0  &0.02800  &NED	&0.173    &1       \\    
1999br  &NGC 4900	  &13 00 41.80  &$+02$ 29 46.0  &0.00320  &NED	&0.024    &2       \\    
1999ca  &NGC 3120         &10 05 22.90  &$-34$ 12 41.0  &0.00931  &NED   &0.109    &2       \\
1999cr  &ESO 576--G034	  &13 20 18.30  &$-20$ 08 50.0  &0.02020  &NED	&0.098    &2       \\    
1999em  &NGC 1637 	  &04 41 27.04  &$-02$ 51 45.2  &0.00267  &NED	&0.040    &2       \\    
1999gi  &NGC 3184 	  &10 18 17.00  &$+41$ 25 28.0  &0.00198  &NED	&0.017    &3       \\    
0210    &MCG +00--03--054 &01 01 16.80  &$-01$ 05 52.0  &0.05140  &NED   &0.036    &4       \\
2002fa  &NEAT J205221.51  &20 52 21.80  &$+02$ 08 42.0  &0.06000  &NED   &0.099    &4       \\
2002gw  &NGC 922 	  &02 25 02.97  &$-24$ 47 50.6  &0.01028  &NED	&0.020    &4       \\    
2002hj  &NPM1G +04.0097   &02 58 09.30  &$+04$ 41 04.0  &0.02360  &NED   &0.115    &4       \\
2002hx  &PGC 23727 	  &08 27 39.43  &$-14$ 47 15.7  &0.03099  &NED	&0.054    &4       \\    
2003B   &NGC 1097 	  &02 46 13.78  &$-30$ 13 45.1  &0.00424  &NED	&0.027    &4 	 \\    
2003E   &MCG--4--12--004  &04 39 10.88  &$-24$ 10 36.5  &0.01490  &J09   &0.048    &4 	 \\    
2003T   &UGC 4864 	  &09 14 11.06  &$+16$ 44 48.0  &0.02791  &NED	&0.031    &4 	 \\    
2003bl  &NGC 5374 	  &13 57 30.65  &$+06$ 05 36.4  &0.01459  &J09   &0.027    &4       \\    
2003bn  &2MASX J10023529  &10 02 35.51  &$-21$ 10 54.5  &0.01277  &NED	&0.065    &4       \\    
2003ci  &UGC 6212 	  &11 10 23.83  &$+04$ 49 35.9  &0.03037  &NED	&0.060    &4       \\    
2003cn  &IC 849 	  &13 07 37.05  &$-00$ 56 49.9  &0.01811  &J09   &0.021    &4       \\    
2003cx  &NEAT J135706.53  &13 57 06.46  &$-17$ 02 22.6  &0.03700  &NED	&0.094    &4       \\    
2003ef  &NGC 4708 	  &12 49 42.25  &$-11$ 05 29.5  &0.01480  &J09   &0.046    &4       \\    
2003fb  &UGC 11522 	  &20 11 50.33  &$+05$ 45 37.6  &0.01754  &J09   &0.183    &4       \\    
2003gd  &M74 		  &01 36 42.65  &$+15$ 44 20.9  &0.00219  &NED	&0.069    &4, 5, 6 \\    
2003hd  &MCG--04--05--010 &01 49 46.31  &$-21$ 54 37.8  &0.03950  &NED	&0.013    &4       \\    
2003hg  &NGC 7771 	  &23 51 24.13  &$+20$ 06 38.3  &0.01427  &NED	&0.074    &4       \\    
2003hk  &NGC 1085 	  &02 46 25.76  &$+03$ 36 32.2  &0.02265  &NED	&0.037    &4       \\    
2003hl  &NGC 772 	  &01 59 21.28  &$+19$ 00 14.5  &0.00825  &NED	&0.073    &4       \\    
2003hn  &NGC 1448 	  &03 44 36.10  &$-44$ 37 49.0  &0.00390  &NED	&0.014    &4       \\    
2003ho  &ESO 235--G58 	  &21 06 30.56  &$-48$ 07 29.9  &0.01438  &NED	&0.039    &4       \\    
2003ip  &UGC 327          &00 33 15.40  &$+07$ 54 18.0  &0.01801  &NED   &0.066    &4       \\
2003iq  &NGC 772 	  &01 59 19.96  &$+18$ 59 42.1  &0.00825  &NED	&0.073    &4       \\    
2004dj  &NGC 2403         &07 37 17.00  &$+65$ 35 58.1  &0.00044  &NED   &0.040    &7       \\
2004et  &NGC 6946         &20 35 25.30  &$+60$ 07 18.0  &0.00016  &NED   &0.342    &8       \\
2005cs  &NGC 5194	  &13 29 53.40  &$+47$ 10 28.0  &0.00154  &NED   &0.035    &9, 10   \\
\hline
\end{tabular}

\tablenotetext{a}{\,Heliocentric host-galaxy redshift.}
  
\tablenotetext{b}{\,Source of host-galaxy redshift: HP02 $\equiv$
  \citet{HaP02}; J09 $\equiv$ \citet{Jon09}; NED $\equiv$ NASA/IPAC
  Extragalactic Database.}

\tablenotetext{c}{\,Galactic reddenings were obtained from \citet{SFD98}.}

\tablerefs{(1)~Cal\'an/Tololo Supernova Program; (2)~SOIRS;
  (3)~\citet{Le02b}; (4)~Carnegie Type II Supernova Survey (CATS);
  (5)~\citet{vDy03}; (6)~\citet{Hen05}; (7)~\citet{Vin06};
  (8)~\citet{Sah06}; (9)~\citet{Pas06}; (10)~\citet{Tsv06}.}

\normalsize
\end{center}
\end{table}

%% file: t31Pars.tex
\begin{table}[p]
\begin{center}
{\scshape \caption{Parameters from the Analytical Fitting to the $V$~Light Curve}\label{TbAnPar}}
\footnotesize
\vspace{2mm}
\begin{tabular}{lcccccccccc}
\hline\hline
SN Name &$a_0$  &$t_{\rm PT}$\mark &$w_0$   &$p_0$  &$m_0$    &$P$     & $Q$    &$R$     &$\chi_{\nu}^2$ \\
        &[mag]  &[JD--2,448,000]  &[day]  &[mag/day] &[mag] &[mag]   &[day]  &[day]    &        \\
\hline
1991al  &1.970  & 519$\pm$3  &4.713  & 0.009  &19.061  &-0.244  & 492.5  & 28.44  & 1.7    \\
1992af  &2.119  & 861.1      &6.080  & 0.024  &18.094  &-2.614  & 784.5  & 90.07  & 1.7    \\
1992am  &1.968  & 947$\pm$5  &9.898  & 0.006  &21.466  & 0.296  & 836.3  & 37.61  & 2.2    \\
1992ba  &1.976  &1014.0      &6.358  & 0.008  &18.033  &-0.081  & 921.8  & 16.35  & 4.1    \\
1993A   &3.923  &1107$\pm$20 &0.108  & 0.009  &24.229  &-0.392  &1009.5  & 87.23  & 9.8    \\
1999br  &1.937  &3396$\pm$20 &1.902  & 0.013  &19.901  &-2.319  & 206.4  &162.85  & 9.6    \\
1999ca  &2.841  &3373.2      &6.227  & 0.011  &19.975  & 1.860  & 268.4  & 31.70  & 1.8    \\
1999cr  &2.014  &3350$\pm$5  &7.939  & 0.008  &20.684  &-0.395  & 294.0  & 39.02  & 2.6    \\
1999em  &1.978  &3590.1      &4.522  & 0.009  &16.297  &-0.403  & 475.3  & 64.56  & 5.3    \\
1999gi  &2.131  &3645.4      &4.891  & 0.006  &17.470  &-1.941  & 514.2  &  6.96  & 2.1    \\
0210    &2.120  &4592$\pm$5  &5.740  & 0.009  &22.768  &-0.601  & 603.7  & 62.48  & 1.6    \\
2002fa  &1.701  &4577$\pm$10 &6.440  & 0.010  &21.624  &-0.906  & 547.8  & 54.02  & 3.1    \\
2002gw  &2.131  &4661.2      &6.151  & 0.008  &19.691  &-0.451  & 570.0  & 57.69  & 2.3    \\
2002hj  &1.968  &4660.3      &5.973  & 0.014  &21.470  & 0.149  & 570.3  & 46.50  & 4.8    \\
2002hx  &1.704  &4658.9      &5.884  & 0.011  &20.802  &-0.532  & 627.2  & 47.42  & 5.5    \\
2003B   &2.130  &4713.8      &6.172  & 0.011  &18.121  &-0.563  & 641.4  & 63.92  & 1.9    \\
2003E   &2.677  &4766$\pm$5  &7.090  & 0.009  &21.563  &-0.542  & 650.4  & 42.14  & 10.0   \\
2003T   &1.871  &4758.7      &3.671  & 0.010  &21.534  &-0.176  & 676.5  & 46.69  & 2.0    \\
2003bl  &3.886  &4805$\pm$5  &3.803  & 0.011  &21.893  &-1.918  & 682.2  &162.52  &  3.7   \\
2003bn  &2.237  &4813.7      &5.660  & 0.007  &20.076  &-0.168  & 726.7  & 25.60  & 3.2    \\
2003ci  &3.105  &4818$\pm$5  &5.204  & 0.012  &22.687  &-0.705  & 783.0  & 59.46  & 2.1    \\
2003cn  &3.524  &4804$\pm$5  &3.770  & 0.019  &22.587  &-0.654  & 689.7  &129.94  & 6.2    \\
2003cx  &1.980  &4828.5      &8.452  & 0.006  &21.519  &-0.492  & 770.8  & 67.70  & 4.6    \\
2003ef  &2.724  &4870$\pm$5  &5.607  & 0.009  &21.157  & 0.120  & 847.4  & 49.27  & 4.6    \\
2003fb  &1.885  &4874.4      &6.755  & 0.010  &21.482  &-0.572  & 811.1  & 50.70  & 3.6    \\
2003gd  &2.439  &4840.9      &2.210  & 0.010  &17.308  & 1.245  & 766.1  & 42.03  &10.6    \\
2003hd  &2.105  &4952.9      &1.475  & 0.012  &21.767  &-0.207  & 892.5  & 53.77  & 5.9    \\
2003hg  &2.684  &4999$\pm$5  &7.800  & 0.009  &21.259  &-0.199  & 914.7  & 19.66  & 3.2    \\
2003hk  &3.097  &4961$\pm$3  &6.296  & 0.011  &21.591  &-0.180  & 930.6  & 21.75  & 6.3    \\
2003hl  &3.376  &5005$\pm$5  &7.809  & 0.008  &20.973  &-0.103  & 892.3  & 39.14  & 9.3    \\
2003hn  &2.010  &4963.7      &3.054  & 0.013  &17.466  &-0.231  & 923.4  & 52.79  & 8.8    \\
2003ho  &2.319  &4920.9      &4.101  & 0.009  &21.703  &-0.159  & 906.2  &  4.43  & 0.2    \\
2003ip  &2.676  &4946$\pm$20 &6.029  & 0.015  &20.321  &-0.531  & 978.4  & 71.97  & 8.6    \\
2003iq  &1.604  &5020$\pm$5  &8.690  & 0.010  &18.116  &-0.326  & 937.8  & 50.85  & 2.3    \\
2004dj  &2.361  &5283$\pm$3  &5.134  & 0.008  &14.703  &-0.323  & 174.7  & 54.01  & 0.9    \\
2004et  &1.850  &5394$\pm$5  &7.439  &-0.004  &15.810  & 1.969  & 237.5  &145.32  & 8.6    \\
2005cs  &2.883  &5666$\pm$10 &4.830  & 0.004  &17.822  &-0.191  & 575.9  & 10.03  & 2.2    \\
\hline
\end{tabular}

\tablenotetext{*}{\,The minimum uncertainty of $t_{\rm PT}$ is set to a
  conservative value of 2~days, and assigned to those SNe for which an
  uncertainty is not quoted. For three cases $t_{\rm PT}$ is manually 
  determined with an uncertainty of 20~days.}

\normalsize
\end{center}
\end{table}

%% file: t32Exts.tex
\begin{table}[p]
\begin{center}
{\scshape \caption{Host-Galaxy Extinctions}\label{Tb2}}
\footnotesize
\vspace{3mm}
\begin{tabular}{l|cc|c|r}
\hline\hline
SN Name      &$A_V$(spec)\tablenotemark{a}  &Subclass  &$A_V$(\nad)\tablenotemark{b} &$A_V(V-I)$\tablenotemark{c}  \\
\hline       
1991al        &0.31(16)     &silver   &0.31(06)     &$-0.17$(21) \\
1992af        &1.24(31)     &coal     &0.17(15)     &$-0.37$(21) \\
1992am        &\nodata      &         &0.00(43)     &$ 0.52$(23) \\
1992ba        &0.43(16)     &silver   &0.00(03)     &$ 0.30$(21) \\
1993A         &0.00(31)     &bronze   &0.00(54)     &$ 0.06$(25) \\
1999br        &0.25(16)     &silver   &0.00(04)     &$ 0.94$(25) \\
1999ca        &0.12(31)     &coal     &0.34(05)     &$ 0.25$(21) \\
1999cr        &0.47(31)     &coal     &0.69(21)     &$ 0.12$(21) \\
1999em        &0.31(16)     &gold     &1.01(05)     &$ 0.24$(21) \\
1999gi        &0.56(16)     &silver   &0.50(08)     &$ 1.02$(21) \\
0210          &0.31(31)     &bronze   &0.00(23)     &$ 0.31$(36) \\
2002fa        &\nodata      &         &0.00(14)     &$-0.35$(26) \\
2002gw        &0.40(19)     &silver   &0.00(02)     &$ 0.18$(22) \\
2002hj        &0.16(31)     &bronze   &0.00(06)     &$ 0.24$(22) \\
2002hx        &0.16(25)     &coal     &0.00(16)     &$ 0.38$(22) \\
2003B\mark    &0.00(25)     &silver   &0.12(05)     &$-0.09$(21) \\ 
2003E         &1.09(31)     &coal     &0.71(07)     &$ 0.78$(23) \\
2003T         &0.53(31)     &coal     &0.19(19)     &$ 0.35$(21) \\
2003bl\mark   &0.00(16)     &gold     &0.11(10)     &$ 0.26$(21) \\
2003bn\mark   &0.09(16)     &silver   &0.00(03)     &$-0.04$(21) \\ 
2003ci        &0.43(31)     &coal     &0.00(23)     &$ 0.78$(23) \\ 
2003cn\mark   &0.00(25)     &gold     &0.00(09)     &$-0.04$(23) \\
2003cx        &0.65(25)     &coal     &\nodata      &$-0.27$(25) \\ 
2003ef        &1.24(25)     &gold     &1.40(12)     &$ 0.98$(21) \\
2003fb        &0.37(31)     &coal     &0.54(23)     &$ 1.24$(23) \\
2003gd        &0.40(31)     &coal     &0.00(04)     &$ 0.33$(21) \\
2003hd        &0.90(31)     &coal     &0.74(27)     &$ 0.01$(21) \\ 
2003hg        &\nodata      &         &2.29(12)     &$ 1.97$(24) \\ 
2003hk        &0.65(31)     &coal     &1.69(20)     &$ 0.44$(25) \\ 
2003hl        &1.24(25)     &gold     &1.84(09)     &$ 1.72$(23) \\ 
2003hn        &0.59(25)     &coal     &0.64(08)     &$ 0.46$(21) \\ 
2003ho        &1.24(31)     &bronze   &1.28(10)     &$ 2.19$(21) \\ 
2003ip        &0.40(31)     &coal     &0.42(08)     &$ 0.56$(22) \\
2003iq        &0.37(16)     &silver   &0.91(04)     &$ 0.25$(22) \\ 
2004dj        &0.50(25)     &coal     &0.26(06)     &$-0.09$(23) \\
2004et        &0.00(25)     &coal     &1.17(02)     &$ 0.13$(27) \\
2005cs        &\nodata      &         &\nodata      &$ 0.72$(30) \\
\hline
\end{tabular}

\tablenotetext{a}{\,Spectroscopic reddenings assuming
  $R_V=3.1$. Additionally, the third column lists the subclass defined
  from the criteria given in \S~\ref{dtc}.}

\tablenotetext{b}{\,Equivalent width of \nad\ measured by us and
  converted to $A_V$ with the relationship of \citet{Bar90} using
  $R_V=3.1$.}

\tablenotetext{c}{\,This paper.}

\tablenotetext{*}{\,SNe selected to determine the intrinsic color.}

\normalsize
\end{center}
\end{table}

%% file: t41HDs.tex
\begin{table}[p]
\begin{center}
{\scshape \caption{Magnitudes, Expansion Velocities, and $V-I$ Colors for Day~$-30$}\label{TbHDs}}
\footnotesize
\vspace{2mm}
\begin{tabular}{lrccccc}
\hline\hline
SN Name &$cz_{\rm CMB}$\mark   &$B$     &$V$           &$I$        &$\upsilon_{\rm Fe~II}$ & $V-I$ \\
        &$\pm187$~[\kmpsec]    &[mag]   &[mag]         &[mag]      &[\kmpsec]               &[mag]  \\
\hline                                                                          
1991al    & 4480  &17.944(036)   &16.917(029)   &16.333(020)    &5328(202)      &0.589(022)  \\
1992af    & 5359  &18.024(038)   &17.096(029)   &16.706(053)    &4529(212)      &0.511(027)  \\
1992am    &14007  &20.443(117)   &19.053(038)   &18.218(033)    &\nodata        &0.830(046)  \\
1992ba    & 1245  &16.962(040)   &15.657(031)   &14.886(022)    &2237(068)      &0.776(021)  \\
1993A     & 8907  &20.636(127)   &19.309(089)   &18.657(057)    &\nodata        &0.678(047)  \\
1999br    & 1285  &19.010(022)   &17.582(013)   &16.574(008)    &1127(205)      &1.028(011)  \\
1999ca    & 3108  &17.901(071)   &16.489(055)   &15.735(034)    &\nodata        &0.755(025)  \\
1999cr    & 6363  &19.435(067)   &18.418(051)   &17.723(035)    &3095(207)      &0.703(025)  \\
1999em    &  670  &15.331(034)   &13.998(023)   &13.245(017)    &2727(056)      &0.752(019)  \\
1999gi    &  831  &16.554(032)   &15.060(019)   &14.011(022)    &2725(061)      &1.058(025)  \\
0210      &15082  &21.955(044)   &20.572(028)   &19.733(035)    &4923(206)      &0.780(110)  \\
2002fa    &17847  &21.222(072)   &20.243(052)   &19.709(088)    &4176(061)      &0.518(065)  \\
2002gw    & 2878  &18.575(020)   &17.491(029)   &16.752(014)    &2669(063)      &0.726(035)  \\
2002hj    & 6869  &19.705(084)   &18.582(061)   &17.939(042)    &3657(065)      &0.751(031)  \\
2002hx    & 9573  &20.328(056)   &19.164(035)   &18.360(028)    &3960(203)      &0.805(031)  \\
2003B     & 1105  &17.078(022)   &15.966(018)   &15.341(015)    &2795(067)      &0.620(017)  \\
2003E     & 4380  &19.659(046)   &18.368(026)   &17.472(024)    &2869(203)      &0.965(024)  \\
2003T     & 8662  &20.541(028)   &19.227(020)   &18.423(022)    &2803(203)      &0.795(024)  \\
2003bl    & 4652  &20.158(042)   &18.961(020)   &18.230(019)    &1475(202)      &0.758(025)  \\
2003bn    & 4173  &18.638(045)   &17.403(034)   &16.769(024)    &2719(202)      &0.641(020)  \\
2003ci    & 9468  &\nodata       &19.634(035)   &18.648(029)    &2876(077)      &0.963(034)  \\
2003cn    & 5753  &19.852(024)   &18.827(015)   &18.182(014)    &2510(207)      &0.640(017)  \\
2003cx    &11282  &20.417(100)   &19.513(050)   &19.050(067)    &3734(205)      &0.551(058)  \\
2003ef    & 4504  &19.304(032)   &17.837(024)   &16.792(017)    &\nodata        &1.043(015)  \\
2003fb    & 4996  &20.521(122)   &19.085(092)   &17.940(056)    &3120(206)      &1.148(040)  \\
2003gd    &  359  &15.208(045)   &13.965(036)   &13.167(023)    &2976(230)      &0.786(019)  \\
2003hd    &11595  &20.462(031)   &19.320(014)   &18.658(018)    &3741(069)      &0.661(022)  \\
2003hg    & 3921  &20.247(091)   &18.067(040)   &16.603(025)    &3398(211)      &1.435(027)  \\
2003hk    & 6568  &19.478(036)   &18.201(024)   &17.380(023)    &3618(087)      &0.828(026)  \\
2003hl    & 2198  &19.124(062)   &17.219(038)   &15.806(027)    &2354(063)      &1.336(027)  \\
2003hn    & 1102  &16.416(020)   &15.153(011)   &14.307(015)    &3121(074)      &0.836(018)  \\
2003ho    & 4134  &20.779(084)   &18.946(026)   &17.419(016)    &4152(080)      &1.523(022)  \\
2003ip    & 5050  &18.894(046)   &17.549(035)   &16.664(025)    &3813(204)      &0.876(021)  \\
2003iq    & 2198  &17.396(049)   &16.124(037)   &15.362(024)    &3482(063)      &0.756(018)  \\
2004dj    &  180  &12.973(074)   &12.046(035)   &11.416(033)    &2725(202)      &0.621(043)  \\
2004et    &$-133$ &13.683(220)   &12.099(171)   &11.378(104)    &2901(202)      &0.707(070)  \\
2005cs    &  635  &16.099(062)   &14.749(051)   &13.824(055)    &\nodata        &0.957(067)  \\
\hline
\end{tabular}

\tablenotetext{*}{\,The {\it Velocity Calculator} tool at the
  NASA/IPAC Extragalactic Database webpage computes the CMB redshift
  ($cz_{\rm CMB}$ in \kmpsec) given the coordinates and the
  heliocentric redshift of an object. Its error corresponds to the
  uncertainty in the determination of the Local Group velocity
  (187~\kmpsec).}

\tablecomments{All of the data in this table have been interpolated to
  day~$-30$ with errors accounting for the interpolation, the error in
  $t_{\rm PT}$, and the instrumental error. $BVI$ corresponds to
  apparent magnitudes corrected for Galactic absorption and
  $K$-terms, but uncorrected for host-galaxy extinction. The $V-I$
  colors do not exactly match the $V$ minus $I$ magnitude difference,
  since light and color curves are interpolated independently.}

\normalsize
\end{center}
\end{table}

%% file: t42Param.tex
\begin{table}[p]
\begin{center}
{\scshape \caption{Fitting Parameters from the Hubble Diagrams}\label{TbPar}}
\vspace{4mm}
\begin{tabular}{l|c|ccc}
\hline\hline
Reddening & & & & \\
estimator                       &filter &{\large $\alpha$} &{\large $\beta$} &{\large $zp$}    \\
\hline			       
$(V-I)$\tablenotemark{a}          &$B$    &$3.27\pm0.37$     &\nodata          &$-0.24\pm0.08$   \\
                                &$V$    &$3.01\pm0.36$     &\nodata          &$-1.36\pm0.08$   \\
                                &$I$    &$2.99\pm0.36$     &\nodata          &$-2.02\pm0.08$   \\
\hline
Spectral                        &$B$    &$4.77\pm0.47$     &\nodata          &$-0.60\pm0.11$   \\
analysis\tablenotemark{b}       &$V$    &$4.22\pm0.45$     &\nodata          &$-1.60\pm0.10$   \\
                                &$I$    &$3.67\pm0.43$     &\nodata          &$-2.11\pm0.10$   \\
\hline
$(V-I)$ +                         &$B$    &$3.50\pm0.30$     &$2.67\pm0.13$    &$-1.99\pm0.11$   \\
variable $R_V$\tablenotemark{c} &$V$    &$3.08\pm0.25$     &$1.67\pm0.10$    &$-2.38\pm0.09$   \\
                                &$I$    &$2.62\pm0.21$     &$0.60\pm0.09$    &$-2.23\pm0.07$   \\
\hline\end{tabular}

\tablenotetext{a}{\,Reddenings from \S~\ref{aho}.}

\tablenotetext{b}{\,Spectroscopic reddenings.}

\tablenotetext{c}{\,Leaving $R_V$ as a free parameter.}

\tablecomments{These are the results of fitting
  Equations~(\ref{HDeq1}) and~(\ref{HDeq2}) to the data in
  Tables~\ref{Tb2} and~\ref{TbHDs}.}

\end{center}
\end{table}

%% file: t43Calib.tex
\begin{table}[p]
\begin{center}
{\scshape \caption{$H_0$ Calculations}\label{TbHo}}
\footnotesize
\vspace{2mm}
\begin{tabular}{l|ccc|r@{$\,\pm\,$}lr@{$\,\pm\,$}lr@{$\,\pm\,$}l}
\hline\hline
        &\multicolumn{3}{c|}{$M+\alpha\log(\upsilon/5000)-\beta(V-I)$\ \ [mag]} &\multicolumn{6}{c}{$H_0$\ \ [\dimho] } \\
\cline{2-10}
SN      &$B$                 &$V$                 &$I$                       &\multicolumn{2}{c}{$H_0(B)$}  &\multicolumn{2}{c}{$H_0(V)$}  &\multicolumn{2}{c}{$H_0(I)$} \\
\hline                                                                       
1999em  &$-17.94\pm0.20$     &$-18.41\pm0.20$     &$-18.24\pm0.19$           &64.7 &6.8                     &62.2  &6.1                    &62.9  &6.0    \\
2004dj  &$-17.08\pm0.30$     &$-17.28\pm0.27$     &$-17.13\pm0.26$           &95.7 &14.0                    &104.5 &13.7                   &104.8 &12.9   \\
\hline                                                                       
Average &$-17.67\pm0.39$     &$-18.02\pm0.53$     &$-17.84\pm0.53$           &{\bf 71} &{\bf 12}            &{\bf 69} &{\bf 16}            &{\bf 70} &{\bf 16} \\
\hline
\end{tabular}

\tablecomments{The uncertainties in $\langle H_0\rangle$ and $\langle
  M_{\rm corr}\rangle$ (the corrected absolute magnitude for $BVI$ in
  the second column) correspond to the weighted dispersion in each
  pair of values, which are a factor of 2--3 greater than the formal
  errors. This is a conservative uncertainty which takes into account
  the larger than expected discrepancy in absolute magnitude between
  the two objects.}

\normalsize
\end{center}
\end{table}

%% file: t44Dist.tex
\begin{table}[p]
\begin{center}
{\scshape \caption{Distance Moduli [mag]}\label{TbMu}}
\small
\vspace{3mm}
\begin{tabular}{lc|ccc|c}
\hline\hline
SN Name  &Host Galaxy     &$\mu_B$       &$\mu_V$      &$\mu_I$     &$\langle\mu\rangle$ \\
\hline
1991al   &LEDA 140858     &34.14(41)  &34.04(54)  &33.89(53)  &34.05(28) \\
1992af  &ESO 340--G038    &34.18(41)  &34.13(54)  &34.13(54)  &34.15(28) \\
1992ba   &NGC 2082        &31.34(40)  &31.31(54)  &31.34(53)  &31.33(28) \\
1999br   &NGC 4900        &31.67(48)  &31.89(59)  &32.10(57)  &31.86(31) \\
1999cr   &ESO 576--G034	  &34.50(42)  &34.63(55)  &34.59(54)  &34.56(28) \\
1999em   &NGC 1637 	  &30.07(40)  &29.96(54)  &29.94(53)  &30.00(28) \\
1999gi   &NGC 3184 	  &30.48(40)  &30.51(54)  &30.52(53)  &30.50(28) \\
0210   &MCG +00--03--054  &37.52(50)  &37.27(57)  &37.09(54)  &37.31(31) \\
2002fa   &NEAT J205221.51 &37.23(44)  &37.16(55)  &37.03(54)  &37.16(29) \\
2002gw   &NGC 922 	  &33.35(41)  &33.46(54)  &33.44(53)  &33.41(28) \\
2002hj   &NPM1G+04.0097   &34.90(41)  &34.93(54)  &34.97(53)  &34.93(28) \\
2002hx   &PGC 23727 	  &35.49(42)  &35.53(54)  &35.45(54)  &35.49(28) \\
2003B    &NGC 1097 	  &32.21(40)  &32.18(54)  &32.15(53)  &32.18(27) \\
2003E  &MCG--4--12--004   &33.91(42)  &34.04(54)  &34.10(54)  &34.01(28) \\
2003T    &UGC 4864 	  &35.21(42)  &35.15(54)  &35.13(54)  &35.17(28) \\
2003bl   &NGC 5374 	  &33.95(45)  &34.09(57)  &34.23(55)  &34.07(30) \\
2003bn   &2MASX J10023529 &33.67(42)  &33.54(55)  &33.53(54)  &33.60(28) \\
2003ci   &UGC 6212 	  &\nodata    &35.31(54)  &35.28(53)  &35.30(38) \\
2003cn   &IC 849 	  &34.77(42)  &34.86(55)  &34.85(54)  &34.81(28) \\
2003cx   &NEAT J135706.53 &36.17(44)  &36.23(55)  &36.23(54)  &36.20(29) \\
2003fb   &UGC 11522 	  &34.41(44)  &34.56(55)  &34.55(54)  &34.49(29) \\
2003gd   &M74 		  &29.99(42)  &29.98(55)  &29.94(54)  &29.98(28) \\
2003hd &MCG--04--05--010  &35.93(40)  &35.85(54)  &35.77(53)  &35.86(28) \\
2003hg   &NGC 7771 	  &33.50(42)  &33.18(54)  &33.14(54)  &33.31(28) \\
2003hk   &NGC 1085 	  &34.45(40)  &34.41(54)  &34.35(53)  &34.41(28) \\
2003hl   &NGC 772 	  &32.08(41)  &32.01(54)  &32.00(53)  &32.04(28) \\
2003hn   &NGC 1448 	  &31.14(40)  &31.15(54)  &31.11(53)  &31.13(27) \\
2003ho   &ESO 235--G58 	  &34.10(41)  &34.18(54)  &34.13(53)  &34.13(28) \\
2003ip   &UGC 327         &33.81(41)  &33.75(54)  &33.67(54)  &33.76(28) \\
2003iq   &NGC 772 	  &32.50(40)  &32.40(54)  &32.34(53)  &32.43(28) \\
2004dj   &NGC 2403        &28.06(43)  &28.22(55)  &28.19(54)  &28.14(29) \\
2004et   &NGC 6946        &28.64(50)  &28.21(58)  &28.17(55)  &28.37(31) \\
\hline
\end{tabular}

\tablecomments{These values were calculated according to the approach
  discussed in \S~\ref{DIST}.  The mean value is a weighted average of
  the individual values for each filter.}

\normalsize
\end{center}
\end{table}

%% file: t45Comp.tex
\begin{table}[p]
\begin{center}
{\scshape \caption{EPM and SCM Distances}\label{TbEPM}}
\vspace{4mm}
\begin{tabular}{lr@{~}lr@{~}lr@{~}l}
\hline\hline
SN Name  &\multicolumn{2}{c}{$d_{\rm E96}$\tablenotemark{a}}  &\multicolumn{2}{c}{$d_{\rm D05}$\tablenotemark{b}}   &\multicolumn{2}{c}{$d_{\rm SCM}$} \\  
         &\multicolumn{2}{c}{[Mpc]} &\multicolumn{2}{c}{[Mpc]} &\multicolumn{2}{c}{[Mpc]}  \\
\hline
1992ba    &16.4 &(2.5)                                  &27.2  &(6.5)                  &18.5  &(2.4)                    \\  
1999br    &\multicolumn{2}{c}{\nodata}                  &39.5  &(13.5)                 &23.6  &(3.4)                    \\  
1999em    &9.3  &(0.5)                                  &13.9  &(1.4)                  &10.0  &(1.3)                    \\  
1999gi    &11.7 &(0.8)                                  &17.4  &(2.3)                  &12.6  &(0.6)                    \\  
2002gw    &37.4 &(4.9)                                  &63.9  &(17.0)                 &48.1  &(6.2)                    \\  
2003T     &87.8 &(13.5)                                 &147.3 &(35.7)                 &108.1 &(14.0)                   \\  
2003bl    &\multicolumn{2}{c}{\nodata}                  &92.4  &(14.2)                 &65.2  &(9.0)                    \\  
2003bn    &50.2 &(7.0)                                  &87.2  &(28.0)                 &52.5  &(6.8)                    \\  
2003hl    &17.7 &(2.1)                                  &30.3  &(6.3)                  &25.6  &(3.3)                    \\  
2003hn    &16.9 &(2.2)                                  &26.3  &(7.1)                  &16.8  &(2.1)                    \\  
2003iq    &36.0 &(5.6)                                  &53.3  &(17.1)                 &30.6  &(4.0)                    \\  
\hline
\end{tabular}

\tablenotetext{a}{\,EPM distances from atmosphere models by
  \citet{Eas96}; see \citet{Jon09}.}

\tablenotetext{b}{\,EPM distances from atmosphere models by
  \citet{DeH05}; see \citet{Jon09}.}

\tablecomments{Uncertainties are given in parentheses.}

\end{center}
\end{table}

%% file: 1SCMv5.1online.bbl
\begin{thebibliography}{}
\bibitem[Allen et al.(2008)]{All08} 
  Allen, S. W., Rapetti, D. A., Schmidt, R. W., Ebeling, H., Morris,
  R. G., \& Fabian, A. C. 2008, MNRAS, 383, 879

\bibitem[Altavilla et al.(2004)]{Alt04}
  Altavilla, G., et al.\ 2004, \mnras, 349, 1344
 
\bibitem[Anderson \& James(2008)]{AnJ08} 
  Anderson, J.~P., \& James, P.~A.\ 2008, \mnras, 390, 1527
 
\bibitem[Arnett(1996)]{Arn96}
  Arnett, D.\ 1996, Supernovae and Nucleosynthesis (New Jersey:
  Princeton Univ. Press)
 
\bibitem[Astier et al.(2006)]{Ast06}
  Astier, P., et al.\ 2006, \aap, 447, 31
 
\bibitem[Baade \& Zwicky(1934)]{BaZ34}
  Baade, W., \& Zwicky, F.\ 1934, Physical Review, 46, 76
 
\bibitem[Barbon et al.(1990)]{Bar90} 
  Barbon, R., Benetti, S., Rosino, L., Cappellaro, E., \& Turatto, M.\
  1990, \aap, 237, 79
 
\bibitem[Bennett et al.(2003)]{Ben03}
  Bennett, C.~L., et al.\ 2003, \apjs, 148, 1
 
\bibitem[Bersten \& Hamuy(2009)]{Ber09}
  Bersten, M.~C., \& Hamuy, M.\ 2009, \apj, 701, 200
 
\bibitem[Blake \& Glazebrook(2003)]{BlG03}
  Blake, C., \& Glazebrook, K.\ 2003, \apj, 594, 665

\bibitem[Blondin et al.(2009)]{Blo09}
  Blondin, S., Prieto, J.~L., Patat, F., Challis, P., Hicken, M.,
  Kirshner, R.~P., Matheson, T., \& Modjaz, M.\ 2009, \apj, 693, 207

\bibitem[Botticella et al.(2008)]{Bot08}
  Botticella, M.~T., et al.\ 2008, \aap, 479, 49
 
\bibitem[Boughn \& Crittenden(2004)]{BoC04}
  Boughn, S., \& Crittenden, R. 2004, Nature, 427, 45
 
\bibitem[Brosch \& Loinger(1991)]{BrL91}
  Brosch, N., \& Loinger, F.\ 1991, \aap, 249, 327
 
\bibitem[Burrows(2000)]{Bur00}
  Burrows, A.\ 2000, \nat, 403, 727
 
\bibitem[Capaccioli et al.(2003)]{Cap03}
  Capaccioli, M., Cappellaro, E., Mancini, D., \& Sedmak, G.\ 2003,
  Memorie della Societa Astronomica Italiana Supplement, 3, 286
 
\bibitem[Cappellaro et al.(1999)]{Cap99}
  Cappellaro, E., Evans, R., \& Turatto, M.\ 1999, \aap, 351, 459
 
\bibitem[Cardelli et al.(1989)]{Car89}
  Cardelli, J.~A., Clayton, G.~C., \& Mathis, J.~S.\ 1989, \apj, 345,
  245
 
\bibitem[Castander(2007)]{Cas07}
  Castander, F.~J.\ 2007, Cosmic Frontiers, 379, 285
 
\bibitem[Clayton \& Cardelli(1988)]{ClC88}
  Clayton, G.~C., \& Cardelli, J.~A.\ 1988, \aj, 96, 695
 
\bibitem[Cousins(1971)]{Cou71}
  Cousins, A.~W.~J.\ 1971, Royal Observatory Annals, 7, C
 
\bibitem[Della Valle \& Panagia(1992)]{DeP92}
  Della Valle, M., \& Panagia, N.\ 1992, \aj, 104, 696
 
\bibitem[Dessart \& Hillier(2005)]{DeH05} 
  Dessart, L., \& Hillier, D.~J.\ 2005, \aap, 439, 671
 
\bibitem[Dessart \& Hillier(2006)]{DeH06}
  Dessart, L., \& Hillier, D.~J.\ 2006, \aap, 447, 691
 
\bibitem[Dessart et al.(2008)]{Des08}
  Dessart, L., et al.\ 2008, \apj, 675, 644
 
\bibitem[Draine(2003)]{Dra03}
  Draine, B.~T.\ 2003, \araa, 41, 241 

\bibitem[Eastman et al.(1996)]{Eas96}
  Eastman, R. G., Schmidt, B. P., \& Kirshner, R. 1996, \apj, 466, 911
 
\bibitem[Emerson et al.(2004)]{Eme04}
  Emerson, J.~P., Sutherland, W.~J., McPherson, A.~M., Craig, S.~C.,
  Dalton, G.~B., \& Ward, A.~K.\ 2004, The Messenger, 117, 27
 
\bibitem[Fabricant et al.(1998)]{Fab98}
  Fabricant, D., Cheimets, P., Caldwell, N., \& Geary, J.\ 1998,
  \pasp, 110, 79
 
\bibitem[Filippenko(1982)]{Fil82}
  Filippenko, A.~V.\ 1982, \pasp, 94, 715
 
\bibitem[Filippenko(1997)]{Fil97}
  Filippenko, A.~V.\ 1997, \araa, 35, 309
 
\bibitem[Finkelman et al.(2008)]{Fin08}
  Finkelman, I., et al.\ 2008, \mnras, 390, 969
 
\bibitem[Fixsen et al.(1996)]{Fix96}
  Fixsen, D.~J., Cheng, E.~S., Gales, J.~M., Mather, J.~C., Shafer,
  R.~A., \& Wright, E.~L.\ 1996, \apj, 473, 576
 
\bibitem[Folatelli et al.(2010)]{Fol10}
  Folatelli, G., et al.\ 2010, \aj, 139, 120
 
\bibitem[Foley et al.(2003)]{F2003}
  Foley, R. J., et al. 2003, \pasp, 115, 1220
 
\bibitem[Fosalba et al.(2003)]{Fos03}
  Fosalba, P., et al. 2003, ApJ, 597, L89
 
\bibitem[Frieman et al.(2008)]{Fri08}
  Frieman, J.~A., Turner, M.~S., \& Huterer, D.\ 2008, \araa, 46, 385
 
\bibitem[Freedman et al.(2001)]{Fre01}
  Freedman, W.~L., et al.\ 2001, \apj, 553, 47
 
\bibitem[Freedman et al.(2009)]{Fre09}
  Freedman, W.~L., et al.\ 2009, \apj, 704, 1036
 
\bibitem[Goobar(2008)]{Goo08}
  Goobar, A.\ 2008, \apjl, 686, L103
 
\bibitem[Granlund et al.(2006)]{Gra06}
  Granlund, A., Conroy, P.~G., Keller, S.~C., Oates, A.~P., Schmidt,
  B., Waterson, M.~F., Kowald, E., \& Dawson, M.~I.\ 2006, \procspie,
  6269, 69
 
\bibitem[Hamuy(2001)]{Ha01b} 
  Hamuy, M.\ 2001, Ph.D. Thesis, Univ.~Arizona
 
\bibitem[Hamuy(2003a)]{Ha03a} 
  Hamuy, M.\ 2003a, ArXiv Astrophysics e-prints,
  arXiv:astro-ph/0301281

\bibitem[Hamuy(2003b)]{Ha03b} 
  Hamuy, M.\ 2003b, ArXiv Astrophysics e-prints,
  arXiv:astro-ph/0309122
 
\bibitem[Hamuy(2005)]{Ham05}
  Hamuy, M.\ 2005, in Cosmic Explosions,
   ed. J. M. Marcaide \& K. W. Weiler (Berlin: Springer-Verlag), 535
 
\bibitem[Hamuy et al.(1996)]{Ha96a}
  Hamuy, M., Phillips, M.~M., Suntzeff, N.~B., Schommer, R.~A., Maza,
  J., \& Aviles, R.\ 1996, \aj, 112, 2391

\bibitem[Hamuy \& Pinto(2002)]{HaP02} 
  Hamuy, M., \& Pinto, P.~A.\ 2002, \apjl, 566, L63
  
\bibitem[Hamuy et al.(1990)]{Ham90}
  Hamuy, M., Suntzeff, N.~B., Bravo, J., \& Phillips, M.~M.\ 1990,
  \pasp, 102, 888
 
\bibitem[Hamuy et al.(1994)]{Ham94}
  Hamuy, M., Suntzeff, N.~B., Heathcote, S.~R., Walker, A.~R., Gigoux,
  P., \& Phillips, M.~M. 1994, \pasp , 106, 566
 
\bibitem[Hamuy et al.(1992)]{Ham92}
  Hamuy, M., Walker, A.~R., Suntzeff, N.~B., Gigoux, P., Heathcote,
  S.~R., \& Phillips, M.~M.\ 1992, \pasp, 104, 533

\bibitem[Hamuy et al.(2001)]{Ha01a}
  Hamuy, M., et al.\ 2001, \apj, 558, 615 
 
\bibitem[Hendry et al.(2005)]{Hen05} 
  Hendry, M.~A., et al.\ 2005, \mnras, 359, 906
 
\bibitem[Hicken et al.(2009)]{Hic09}
  Hicken, M., Wood-Vasey, W.~M., Blondin, S., Challis, P., Jha, S.,
  Kelly, P.~L., Rest, A., \& Kirshner, R.~P.\ 2009, \apj, 700, 1097
 
\bibitem[Hobbs (1974)]{Hob74}
  Hobbs, L.~M., 1974, \apj, 191, 381
 
\bibitem[Hodapp et al.(2004)]{Hod04}
  Hodapp, K.~W., et al.\ 2004, Astronomische Nachrichten, 325, 636 

\bibitem[Hopkins \& Beacom(2006)]{Hop06}
  Hopkins, A.~M., \& Beacom, J.~F.\ 2006, \apj, 651, 142
 
\bibitem[Janka et al.(2007)]{Jan07}
  Janka, H.-T., Langanke, K., Marek, A., Mart\'inez-Pinedo, G.,
  M\"uller, B.\ 2007, \physrep, 442, 38
 
\bibitem[Johnson et al.(1966)]{Joh66}
  Johnson, H.~L., Iriarte, B., Mitchell, R.~I., \& Wisniewski, W.~Z.\
  1966, Commun. Lunar Plan. Lab., 4, 99
 
\bibitem[Jones et al.(2009)]{Jon09}
  Jones, M.~I., et al.\ 2009, \apj, 696, 1176

\bibitem[Kasen \& Woosley(2009)]{KaW09}
  Kasen, D., \& Woosley, S.~E.\ 2009, \apj, 703, 2205

\bibitem[Kirshner \& Kwan(1974)]{KiK74} 
  Kirshner, R.~P., \& Kwan, J.\ 1974, \apj, 193, 27
 
\bibitem[Krisciunas et al.(2009)]{Kri09}
  Krisciunas, K., et al.\ 2009, \aj, 137, 34

\bibitem[Law et al.(2009)]{Law09}
  Law, N.~M., et al.\ 2009, \pasp, 121, 1395
 
\bibitem[Landolt(1992)]{Lan92}
  Landolt, A.~U.\ 1992, \aj, 104, 340
 
\bibitem[Leonard et al.(2002a)]{Le02a} 
  Leonard, D.~C., et al.\ 2002a, \pasp, 114, 35
 
\bibitem[Leonard et al.(2002b)]{Le02b} 
  Leonard, D.~C., et al.\ 2002b, \aj, 124, 2490
 
\bibitem[Nadyozhin(2003)]{Nad03}
  Nadyozhin, D.~K.\ 2003, \mnras, 346, 97
 
\bibitem[Matheson et al.(2008)]{Mat08}
  Matheson, T., et al.\ 2008, \aj, 135, 1598
 
\bibitem[Miller \& Stone(1993)]{Mil93}
  Miller, J. S., \& Stone, R. P. S. 1993, Lick Obs. Tech. Rep. 66
  (Santa Cruz: UCSC)
 
\bibitem[Minkowski(1941)]{Min41} 
  Minkowski, R.\ 1941, \pasp, 53, 224
 
\bibitem[Munari \& Zwitter(1997)]{MuZ97}
  Munari, U., \& Zwitter, T.\ 1997, \aj, 318, 269
 
\bibitem[Nelder \& Mead(1965)]{NeM65} 
  Nelder, J.~A., \& Mead, R.\ 1965, Computer Journal, vol.~7, 308, 313
 
\bibitem[Nugent et al.(2006)]{Nug06} 
  Nugent, P., et al.\ 2006, \apj, 645, 841
 
\bibitem[Olivares(2008)]{Oli08} 
  Olivares, F.\ 2008, M.Sc. Thesis, Univ. de Chile, arXiv:0810.5518
 
\bibitem[Pastorello et al.(2006)]{Pas06} 
  Pastorello, A., et al.\ 2006, \mnras, 370, 1752
 
\bibitem[Perlmutter et al.(1999)]{Per99}
  Perlmutter, S., et al.\ 1999, \apj, 517, 565
 
\bibitem[Phillips(1993)]{Phi93}
  Phillips, M.~M.\ 1993, \apjl, 413, L105
 
\bibitem[Phillips et al.(1999)]{Phi99}
  Phillips, M.~M., Lira, P., Suntzeff, N.~B., Schommer, R.~A., Hamuy,
  M., \& Maza, J.\ 1999, \aj, 118, 1766
 
\bibitem[Poznanski et al.(2009)]{Poz09}
  Poznanski, D., et al.\ 2009, \apj, 694, 1067

\bibitem[Rau et al.(2009)]{Rau09}
  Rau, A., et al.\ 2009, \pasp, 121, 1334
 
\bibitem[Reindl et al.(2005)]{Rei05}
  Reindl, B., Tammann, G.~A., Sandage, A., \& Saha, A.\ 2005, \apj,
  624, 532
 
\bibitem[Riess et al.(1996)]{RPK96}
  Riess, A. G., Press, W. H., \& Kirshner, R. P. 1996, ApJ, 473, 88
 
\bibitem[Riess et al.(1998)]{Rie98}
  Riess, A.~G., et al.\ 1998, \aj, 116, 1009
 
\bibitem[Rifatto(1990)]{Rif90}
  Rifatto, A.\ 1990, in Dusty Objects in The Universe, ed.
  E. Bussoletti \& A.~A. Vittone (Dordrecht: Kluwer), 277
 
\bibitem[Sahu et al.(2006)]{Sah06} 
  Sahu, D.~K., Anupama, G.~C., Srividya, S., \& Muneer, S.\ 2006,
  \mnras, 372, 1315
 
\bibitem[Seo \& Eisenstein(2003)]{SeE03}
  Seo, H.-J., \& Eisenstein, D.~J.\ 2003, \apj, 598, 720
 
\bibitem[Schlegel et al.(1998)]{SFD98} 
  Schlegel, D.~J., Finkbeiner, D.~P., \& Davis, M.\ 1998, \apj, 500,
  525
 
\bibitem[Schmidt et al.(1994)]{Sch94}
  Schmidt, B.~P., Kirshner, R.~P., Eastman, R.~G., Phillips, M.~M.,
  Suntzeff, N.~B., Hamuy, M., Maza, J., \& Aviles, R.\ 1994, \apj,
  432, 42
 
\bibitem[Spergel et al.(2007)]{Spe07}
  Spergel, D.~N., et al.\ 2007, \apjs, 170, 377
 
\bibitem[Tonry et al.(2000)]{Ton00}
  Tonry, J.~L., Blakeslee, J.~P., Ajhar, E.~A., \& Dressler, A.\ 2000,
  \apj, 530, 625

\bibitem[Tonry et al.(2001)]{Ton01}
  Tonry, J.~L., Dressler, A., Blakeslee, J.~P., Ajhar, E.~A.,
  Fletcher, A.~B., Luppino, G.~A., Metzger, M.~R., \& Moore, C.~B.\
  2001, \apj, 546, 681
 
\bibitem[Tripp(1998)]{Tri98}
  Tripp, R.\ 1998, \aap, 331, 815 

\bibitem[Tsvetkov et al.(2006)]{Tsv06} 
  Tsvetkov, D.~Y., Volnova, A.~A., Shulga, A.~P., Korotkiy, S.~A.,
  Elmhamdi, A., Danziger, I.~J., \& Ereshko, M.~V.\ 2006, \aap, 460,
  769

\bibitem[Turatto et al.(2003)]{Tur03}
  Turatto, M., Benetti, S., \& Cappellaro, E.\ 2003, From Twilight to
  Highlight: The Physics of Supernovae, 200

\bibitem[Tyson et al.(2003)]{Tys03}
  Tyson, J.~A., Wittman, D.~M., Hennawi, J.~F., \& Spergel, D.~N.\
  2003, Nuclear Physics B Proceedings Supplements, 124, 21
 
\bibitem[Utrobin(2007)]{Utr07}
  Utrobin, V.~P.\ 2007, \aap, 461, 233
 
\bibitem[Van Dyk et al.(2003)]{vDy03} 
  Van Dyk, S.~D., Li, W., \& Filippenko, A.~V.\ 2003, \pasp, 115, 1289
 
\bibitem[Vink\'o et al.(2006)]{Vin06} 
  Vink\'o, J., et al.\ 2006, \mnras, 369, 1780
 
\bibitem[Wang (2005)]{Wan05}
  Wang, L.\ 2005, \apjl, 635, L33
 
\bibitem[Wang et al.(2009)]{Wan09} 
  Wang, X., et al.\ 2009, \apj, 699, L139
 
\bibitem[Weaver \& Woosley(1980)]{WeW80} 
  Weaver, T.~A., \& Woosley, S.~E.\ 1980, in Ninth Texas Symposium on
  Relativistic Astrophysics, ed. J. Ehlers, J.~J. Perry, \& M. Walker
  (New York: New York Academy of Sciences), 335

\bibitem[Wood-Vasey et al.(2007)]{Woo07}
  Wood-Vasey, W.~M., et al.\ 2007, \apj, 666, 694


\end{thebibliography}
